\documentclass[galaxies,accept,moreauthors,pdftex]{mdpi} 
\usepackage{amssymb}

\firstpage{1} 
\pubvolume{xx}
\issuenum{1}
\articlenumber{5}
\pubyear{2019}
\copyrightyear{2019}
\history{Received: date; Accepted: date; Published: date}

\newcommand\aj{\ref@jnl{AJ}}
\newcommand\araa{\ref@jnl{ARA\&A}}
\newcommand\apj{\ref@jnl{ApJ}}
\newcommand\apjl{\ref@jnl{ApJL}}     
\newcommand\apjs{\ref@jnl{ApJS}}
\newcommand\ao{\ref@jnl{ApOpt}}
\newcommand\apss{\ref@jnl{Ap\&SS}}
\newcommand\aap{\ref@jnl{A\&A}}
\newcommand\aapr{\ref@jnl{A\&A~Rv}}
\newcommand\aaps{\ref@jnl{A\&AS}}
\newcommand\azh{\ref@jnl{AZh}}
\newcommand\baas{\ref@jnl{BAAS}}
\newcommand\icarus{\ref@jnl{Icarus}}
\newcommand\jrasc{\ref@jnl{JRASC}}
\newcommand\memras{\ref@jnl{MmRAS}}
\newcommand\mnras{\ref@jnl{MNRAS}}
\newcommand\pra{\ref@jnl{PhRvA}}
\newcommand\prb{\ref@jnl{PhRvB}}
\newcommand\prc{\ref@jnl{PhRvC}}
\newcommand\prd{\ref@jnl{PhRvD}}
\newcommand\pre{\ref@jnl{PhRvE}}
\newcommand\prl{\ref@jnl{PhRvL}}
\newcommand\pasp{\ref@jnl{PASP}}
\newcommand\pasj{\ref@jnl{PASJ}}
\newcommand\qjras{\ref@jnl{QJRAS}}
\newcommand\skytel{\ref@jnl{S\&T}}
\newcommand\solphys{\ref@jnl{SoPh}}
\newcommand\sovast{\ref@jnl{Soviet~Ast.}}
\newcommand\ssr{\ref@jnl{SSRv}}
\newcommand\zap{\ref@jnl{ZA}}
\newcommand\nat{\ref@jnl{Nature}}
\newcommand\iaucirc{\ref@jnl{IAUC}}
\newcommand\aplett{\ref@jnl{Astrophys.~Lett.}}
\newcommand\apspr{\ref@jnl{Astrophys.~Space~Phys.~Res.}}
\newcommand\bain{\ref@jnl{BAN}}
\newcommand\fcp{\ref@jnl{FCPh}}
\newcommand\gca{\ref@jnl{GeoCoA}}
\newcommand\grl{\ref@jnl{Geophys.~Res.~Lett.}}
\newcommand\jcp{\ref@jnl{JChPh}}
\newcommand\jgr{\ref@jnl{J.~Geophys.~Res.}}
\newcommand\jqsrt{\ref@jnl{JQSRT}}
\newcommand\memsai{\ref@jnl{MmSAI}}
\newcommand\nphysa{\ref@jnl{NuPhA}}
\newcommand\physrep{\ref@jnl{PhR}}
\newcommand\physscr{\ref@jnl{PhyS}}
\newcommand\planss{\ref@jnl{Planet.~Space~Sci.}}
\newcommand\procspie{\ref@jnl{Proc.~SPIE}}

\newcommand\actaa{\ref@jnl{AcA}}
\newcommand\caa{\ref@jnl{ChA\&A}}
\newcommand\cjaa{\ref@jnl{ChJA\&A}}
\newcommand\jcap{\ref@jnl{JCAP}}
\newcommand\na{\ref@jnl{NewA}}
\newcommand\nar{\ref@jnl{NewAR}}
\newcommand\pasa{\ref@jnl{PASA}}
\newcommand\rmxaa{\ref@jnl{RMxAA}}

\newcommand\maps{\ref@jnl{M\&PS}}
\newcommand\aas{\ref@jnl{AAS Meeting Abstracts}}
\newcommand\dps{\ref@jnl{AAS/DPS Meeting Abstracts}}

\Title{The UV Perspective of Low-Mass Star Formation}

\Author{P. C. Schneider$^{1,\dagger \star}$\orcidA{}, H. M. G\"unther$^{2}$\orcidB{}, K. France$^{3}$ \orcidC{}} 

\AuthorNames{Christian Schneider}

\address{%
$^{1}$ \quad Hamburger Sternwarte; astro@pcschneider.eu\\
$^{2}$ \quad Massachusetts Institute of Technology, Kavli Institute for Astrophysics and Space Research\\
$^{3}$ \quad Department of Astrophysical and Planetary Sciences Laboratory for Atmospheric and Space Physics, University of Colorado}

\corres{Correspondence: astro@pcschneider.eu (P. C. S.)}

\firstnote{Current address: Hamburger Sternwarte, Gojenbergsweg 112, 
21029 Hamburg, Germany} 

\abstract{The formation of low-mass ($M_\star\lesssim2\,M_\odot$) stars in molecular
clouds involves accretion disks and jets, which are 
of broad astrophysical interest.  Accreting stars represent the closest
examples of these phenomena. Star and planet formation are also intimately 
connected, setting the starting point for planetary systems like our own. The
ultraviolet (UV) spectral range
is particularly suited to study star formation, because virtually all relevant
processes radiate at temperatures associated with UV emission processes 
or have strong observational signatures in the UV.
In this review, we describe how UV observations provide unique diagnostics for
the accretion process, the physical properties of the protoplanetary disk,
and jets and outflows. 
}


\begin{document}


\section{Introduction}
Stars form in molecular clouds. When these clouds fragment, localized 
cloud regions
collapse into groups of protostars. Stars with final masses between 
$0.08\,M_\odot$ and $2\,M_\odot$, broadly the progenitors of Sun-like stars,
start as cores deeply embedded 
in a dusty envelope, where they can be seen only in the sub-mm and far-IR 
spectral window (so-called class~0 sources). The collapse into a central 
condensation is not spherically symmetric due to angular 
momentum conservation. Instead, an accretion disk forms 
while the remaining envelope still hides the inner core from view in all 
but the mid/far-IR (class~I sources). These young objects also typically
drive powerful jets that propagate through the envelope, pierce the 
cloud, and are often the earliest signs of a forming star. After some time,
the envelope dissipates and the stars become visible in the optical. This marks the transition to class~II sources, also called classical T~Tauri stars (CTTSs). 
The period of time between disk formation and the end of the CTTS phase is 
highly interesting as it is the time when planet formation is thought to take place.
Afterwards,
the circumstellar disk disperses, planet formation halts, and accretion ceases,
typically within a few Myrs. Stars in this stage are called class~III sources 
or weak-line T~Tauri stars (WTTSs). They move along Hayashi tracks
towards the main sequence where they will stay most of their lifetime.
If we want to understand star and planet formation, including the Sun and the 
solar system, ``we must unravel all the mysteries of [CTTSs]'' \citep{Imhoff_1977}.
Of the different evolutionary steps, the CTTS stage is the 
earliest evolutionary step where UV observations can provide detailed information 
about the physical conditions in the disk, mass accretion onto the forming 
protostar, and outflow activity central to the star-formation process.

\subsection{A Short History of CTTSs}

CTTSs were initially identified as a new class of objects 
due their optical variability. Very early on, it was realized that 
they also show signs of enhanced chromospheric activity, i.e., ``emission lines 
resembling those of the solar chromosphere'' \citep{Joy_1945}. These emission
lines are superposed on a photospheric spectrum  similar to main sequence stars of late spectral type.
Initially, the source of emission in excess of a 
normal photosphere was speculated to be circumstellar or chromospheric in origin \citep{Greenstein_1950}.

The discovery that CTTSs are above and to the right of the main sequence in an HR-diagram demonstrated
in conjunction with theoretical work \citep{Henyey_1955},
that these objects must be 
young and, thus, the progenitors of the large main sequence population 
\citep{Ambartsumian_1954,Walker_1956}. This notion is corroborated by the presence
of strong Li absorption, which requires $\sim100\times$ the Li abundance of the 
Sun \citep[e.g.][]{Magazzu_1992}. Because Li is depleted quickly in stellar interiors, the amount of surface 
lithium decreases with time in convective stars. Therefore, strong Li absorption can only be found in young stars and thus CTTS must be young \citep[e.g. review by][]{Pinsonneault_1997}.

Meanwhile the number of peculiarities found in CTTSs grew, but their
established youth alone was insufficient to explain those peculiarities and
the ``mysteries'' of CTTSs remained. Many features, like the origin of the 
``chromospheric emission'' remained unexplained, 
until the early 1980s when a large variety of models were proposed to 
explain features of CTTSs including envelopes, outflows, 
infall, and dust disks. 

Specifically, the ``ultraviolet excess'' (measured
around 3700\,\AA{}) and the ``blue continuum'', which go along with a
weakening of photosperic absorption lines due to veiling 
\citep{Kuhi_1974,Strom_1975}, were shown to 
 correlate with the strength of H$\alpha$ 
\citep{Kuhi_1966, Kuhi_1968}. Such a correlation indicates a common origin of
both features as observed for chromospheric emission on the Sun. Thus, it was
natural to ascribe the excess continuum emission at short wavelengths  to a 
Balmer continuum, i.e.,  emission following the capture of a free electron by 
an ionized hydrogen atom. While that finding pinpoints neither the physical origin
nor location of the emitting material, it paved the way for 
explanations involving a suitably tuned chromosphere for producing the
observed (Balmer \& Ca~II H+K) line fluxes in CTTSs. Quantitative models of deep 
lying chromospheres by \citet{Cram_1979} using ad~hoc temperature profiles, so-called 
deep chromosphere models,  
were able to reproduce the fluxes except for H$\alpha$.


These static chromospheric models, however, had difficulties in explaining 
the kinematic line profiles observed in CTTSs \citep{Dumont_1973} and 
additional emitting regions were required \citep{Brown_1984, Calvet_1984} to explain 
the suite of line properties and overall flux characteristic. In fact, the 
existence of infalling or outflowing 
material was already postulated by \citet{Walker_1963} in 1963. The emission lines observed 
in CTTSs often show P~Cygni profiles that were initially interpreted 
in terms of a hot ($\sim10^4$\,K) outflow \citep{Kuhi_1964}---somewhat 
contradictory to the general infall during the cloud contraction phase  
\citep{Larson_1973}. Interestingly, the mass-loss rates that go along with this 
shell expanding at around 300\,km\,s$^{-1}$ were estimated to a 
few times $10^{-8}\,M_\odot$\,yr$^{-1}$, similar to modern day mass loss estimates (though not in the form of a
spherical mass-loss).

The outflow explanation, however, was not entirely satisfactory, although 
the presence of P~Cygni profiles was thought to conclusively demonstrate
the existence of mass-loss in CTTSs.
Notwithstanding the question why CTTSs should be surrounded by a shell of hot
 gas, the  
inverse P~Cygni profiles expected for mass-infall are mainly seen 
in YY~Ori stars, a class of objects that share many similarities with CTTSs. 
The notion that mass accretion onto the central star also plays a key role 
in CTTSs was proposed when it was discovered that the line 
profiles can cycle between the inverse P~Cygni profile of
``typical'' CTTSs and the inverse P~Cygni profile of YY~Ori objects 
within a few days \citep{Walker_1972, Strom_1975} and it was suggested
that a flattened disk, by analogy with the solar system, exists around 
CTTSs. In fact, the 
IR excess of CTTSs was soon to be found with the advent of IR instruments and 
correctly ascribed to relatively cold ($T\sim 700$\,
K) dust; the alternative of an optically 
thin gaseous envelope would produce booming Balmer continuum emission that is not 
observed \citep{Cohen_1979}. This dust, or rather the flared protoplanetary disk 
around the star, is of exceptional importance for the study of CTTSs. 
In this article, however, we treat the disk just as a reservoir of material 
close to the central star and refer 
the interested reader to recent reviews for more details on protoplanetary disks 
\citep{Dullemond_2010, Williams_2011}.

The disk hypothesis has the great property of leaving a large fraction of 
the stellar photosphere with its absorption lines visible, provided that the 
``ultraviolet excess'' emission is roughly comparable in luminosity. 
\citet{Walker_1972}  postulated a shock forming at the interface
between material falling onto the star and the stellar photosphere as the 
source of the continuum emission and the disk as the source of the Balmer line 
emission---the first mention of the CTTS 
accretion paradigm, which is still largely accepted. Notably, it is still 
believed that the Balmer lines do not originate in the same region as the 
continuum emission although the disk itself is not considered the dominant
emission region anymore. The disk accretion scenario was subsequently put on 
solid theoretical grounds by \citet{Lynden-Bell_1974} 
and  \citet{Shakura_1973}.
The accretion scenario was then specifically applied to CTTSs for explaining
the line profiles including blue-shifted absorption \citep{Ulrich_1976}.  
\citet{Ulrich_1976} already noted that soft X-ray emission from the 
post-shock plasma is expected.

\subsection{The First FUV Observations of Young, Accreting Stars}

\begin{figure}[t!]
\begin{center}
\includegraphics[width=16 cm]{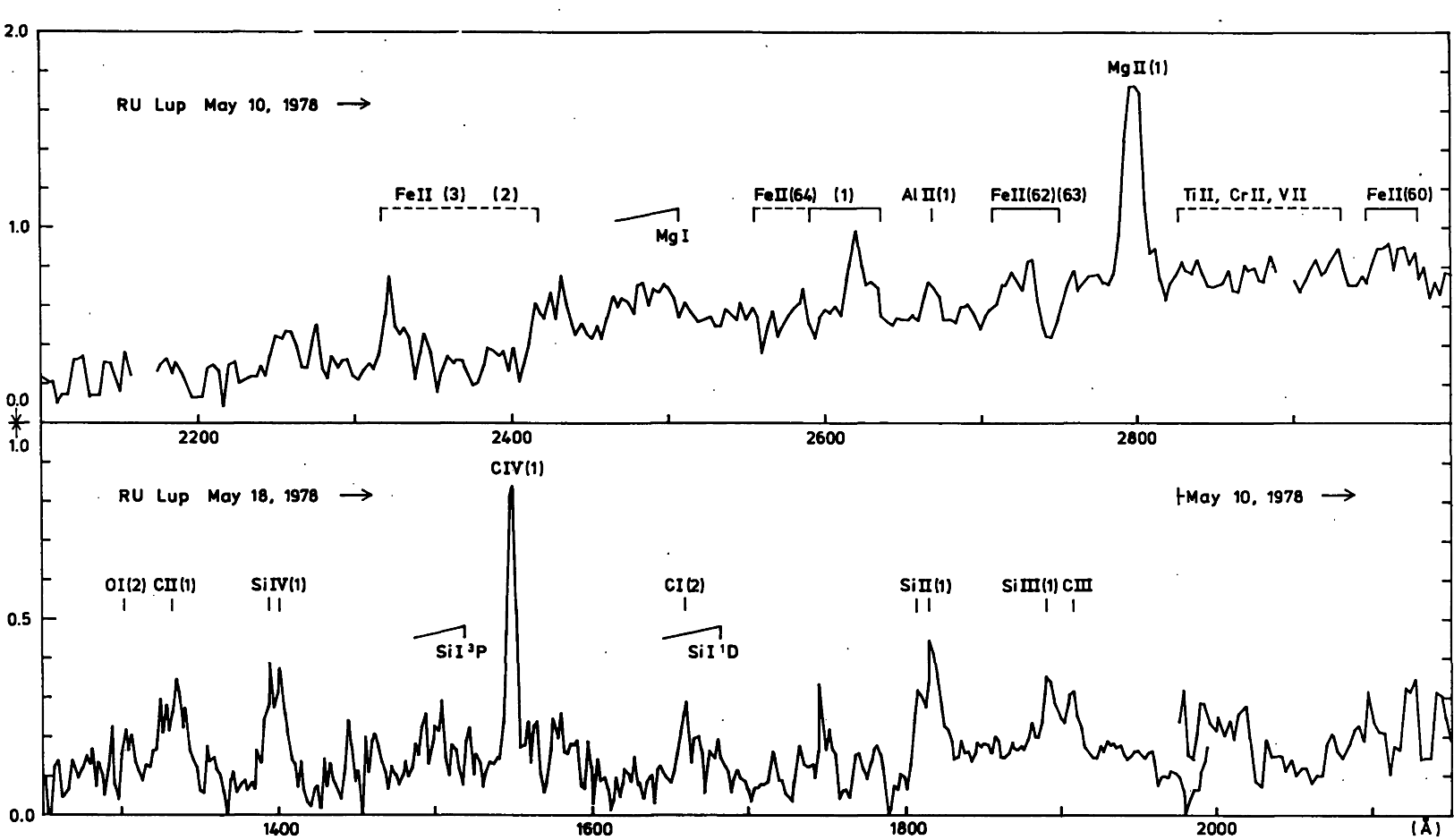}
\end{center}
\caption{IUE spectrum of RU~Lup (flux unit:
     $10^{-13}\,$erg\,cm$^{-2}$\,s$^{-1}$\,\AA$^{-1}$) with the main 
     features labeled, one of the first high-fidelity 
     FUV spectra of a CTTS. From \citet{Gahm_1979}. 
     Reproduced with permission \textcopyright{} ESO. \label{fig:IUE_RU_Lup}}
\end{figure}   

With protoplanetary disks, accretion, 
and outflows in the form of winds and jets recognized as significant processes 
for CTTSs, essentially all major features of CTTS systems were identified 
\cite{Bertout_2007} and required observational support/constraints with the
``\dots ultraviolet spectral region [holding] promise in
helping to resolve many of the issues connected with
the T~Tauri stars.'' \citep{Imhoff_1980}. This statement still holds as 
demonstrated, e.g., by the inception of a Hubble Ultraviolet Legacy program
using HST Director's Discretionary time, the ``UV Legacy Library of Young Stars as Essential Standards'' (ULLYSES) survey, which is anticipated to dedicate $\sim$~500 HST orbits to study
low-mass star formation in the ultraviolet. 

Three features make the 
UV spectral range particularly interesting to study CTTS: (a) Emission from hot 
plasma related to the accretion shock should emit most of its energy in the FUV
range (already mentioned in \citep{Appenzeller_1979}), (b) disk features seen as
fluorescently excited H$_2$ and CO emission \citep{Herczeg_2002, Schindhelm_2012}
and in absorption against the star \citep{France_2011}, and (c) bright atomic 
emission lines from the hot post-shock region of protostellar jets and 
fluorescent emission from wide-angle outflows \citep{Schneider_2013}.

https://www.overleaf.com/project/5e42c54e77c525000110cf18

The first FUV (1550, 1800\,\AA) to NUV (2200, 2500, 3300\,\AA) photometric data 
of CTTSs were obtained by the ultraviolet experiment onboard
the Astronomical Netherlands 
Satellite (ANS, \cite{Duinen_1975}). Targeting seven of the ``most prominent'' 
CTTSs, three were detected \cite{Boer_1977}. In the discovery paper, \citet{Boer_1977} 
specifically discuss the FUV properties of V\,380~Ori, which showed 
excess emission that they associate with the Balmer continuum produced 
by dense gas. Interestingly, these authors already state that dereddening 
FUV fluxes of CTTSs is subject to comparably large  uncertainties, still one of 
the unsolved uncertainties when studying CTTSs (see discussion in 
\citet{France_2017}).

A big step forward for FUV studies of CTTSs was the launch of the IUE satellite 
\cite{Boggess_1978}, which provided the opportunity to measure 
individual lines in larger samples of CTTSs. Figure~\ref{fig:IUE_RU_Lup} shows
an IUE spectrum of the CTTS RU~Lup \cite{Gahm_1979}, still representative
for most CTTSs FUV spectra. The spectrum is dominated by strong emission lines 
from ionized species with the presence of C~{\sc iv}, Si~{\sc iv}, and 
Mg~{\sc ii} and \citet{Gahm_1979} suggested regions with $\log\,T\approx4.5-5.0$ 
``around'' the star, a notion that essentially holds today as most of the hot 
lines are thought to arise from material somehow influenced by accretion.

\begin{figure}[t!]
\begin{center}
\includegraphics[width=0.65\textwidth, angle=270]{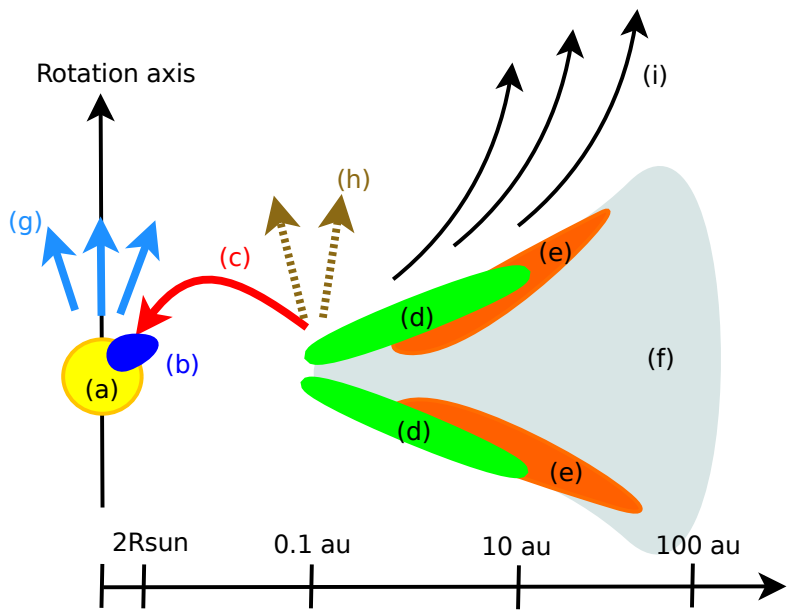}
\end{center}
\caption{Sketch of a CTTS system with regions that are relevant for FUV 
studies labeled. Relative positions are correct, but relative distances 
and sizes are not. The relevant components are (a) the pre-main sequence 
star, (b) the accretion shock on the stellar surface, (c) the accretion funnel, 
(d) the inner, hot disk region, ($T\sim2,000$\,K) (e) cooler regions of the 
inner disk surface ($T\sim500$\,K), 
(f) cool protoplanetary disk material, (g) a spherical (stellar) wind, which may 
be seen in emission or as an absorption component in strong emission 
lines (e.g., C~{\sc iv}), (h) an 
X-wind, (i) a disk wind. 
\label{fig:sketch}}
\end{figure}

Figure~\ref{fig:sketch} provides a sketch of the most relevant FUV emission
regions in CTTSs. In the next sections, we successively describe how FUV 
studies continue to provide key information for the physical processes in CTTSs.
We begin with the intrinsic FUV emission of the star due to enhanced magnetic
activity, move then successively outwards,
i.e., we then discuss the emission thought to arise from accretion, the disk 
features, and emission from outflows and jets. We close with a quick prospectus 
on future UV disk studies.

\section{Stellar UV emission of CTTSs from magnetic activity} 
Young stars are rapid rotators with typical rotation periods of a few days \cite{Bouvier_1986,Bouvier_1993,Herbst_2007, Bouvier_2014}, because 
the stars retain some of the angular momentum from their natal cloud during 
the contraction phase. In addition,
T~Tauri stars possess largely convective interiors while moving from the birth 
line towards the zero age main 
sequence (ZAMS) so that they are able to generate kiloGauss magnetic fields \citep{Gregory_2010}. As a result, T~Tauri stars show all
signs of enhanced magnetic activity in the form of strong chromospheric and coronal 
emission compared to their main sequence siblings \citep{Houdebine_1996,
Feigelson_1999}.

In non-accreting 
stars, chromospheric activity causes emission in excess to that expected from 
the stellar photosphere and particularly WTTSs are highly magnetically active 
\cite{Walter_1987}. This enhanced magnetic activity is due to extended outer, 
rarefied layers of the atmosphere with temperatures between about $T=10^4$ and 
$10^7$\,K. Above the photosphere, one finds the chromosphere ($T\sim10^4$) and a 
transition region (TR, $T\sim10^5$\,K) connecting the chromosphere with the 
corona ($T\sim10^{6-7}$\,K). As a rule of thumb, chromospheric and TR emission 
is most prominent in the UV range while the corona emits mostly in the X-ray 
regime.

For the study of genuine star formation related processes, e.g., accretion, the enhanced
magnetic activity of T~Tauri stars is a nuisance, because the chromospheric and TR emission from magnetic activity must be carefully 
separated/subtracted.  WTTSs are excellent templates for the emission caused by magnetic 
activity, because they lack accretion and gas-rich disks, but share internal structure
and rotation with the CTTSs. Using a magnetically inactive dwarf as a
template for the UV emission instead of a WTTS may result in the erroneous conclusion that 
a particular star is accreting at a significant rate 
($\dot{M} \sim 10^{-8.5}\,M_\odot$\,yr$^{-1}$, \cite{Ingleby_2011}).

\begin{figure}[t!]
\begin{center}
\includegraphics[width=14 cm]{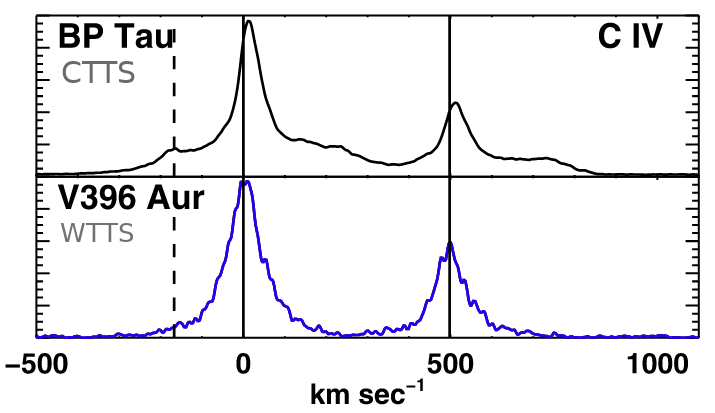}
\end{center}
\caption{Comparison of a typical CTTSs line profile (top), here C~{\sc iv}, and that 
of a comparable mass WTTSs (bottom). The dashed line marks the location of emission from fluorescent H$_{2}$.  From \citet{Ardila_2013}.
\textcopyright{}AAS. Reproduced with permission. \label{fig:BP_Tau_Civ}}
\end{figure}

It is therefore not surprising that many large observing programs studying CTTSs included a number of WTTSs as templates to 
distinguish emission from magnetic activity and, e.g., accretion 
\citep[][and Fig.~\ref{fig:BP_Tau_Civ}]{Costa_2000}. Any emission in excess of the chromospheric to 
coronal flux of WTTSs is, as a starting hypothesis,  thought 
to be attributable to star formation processes such as accretion, disks, or jets.
(In fact, no theory exists that covers self-consistently all FUV emission 
components of CTTSs.) Therefore, we define the term ``excess emission'' 
in the context of this article as the emission not seen in a comparable 
WTTSs although we note that WTTSs show some spread in their 
UV properties \cite{Findeisen_2011}.

\section{Accretion} 
CTTSs possess strong stellar magnetic fields in the kG range \citep[][]{JK_2007}, 
compared to just a few G average surface fields on solar mass main-sequence stars. 
The strong magnetic fields truncate the 
disk at a few stellar radii so that accretion proceeds along magnetic
field lines connecting the star and the inner disk \cite{Koenigl_1991}.
This leads to a
strong shock, where the infalling material impacts the stellar
atmosphere (see Fig.~\ref{fig:sketch}).
Since typical free-fall velocities are around 300\,km\,s$^{-1}$, the post shock 
temperature is in the MK range and the plasma cools predominately through
X-ray emission, which is then partly reprocessed into UV and optical radiation 
\citep[][]{Calvet_1998, Schneider_2017}. The accretion hot spot covers on the 
order of 1\,\% of the stellar surface, so that the densities in 
the accretion funnel should be around $10^{12}$\,cm$^{-3}$ to provide the accretion 
rates in the observed $10^{-8}\,M_\odot$\,yr$^{-1}$ range \cite{Calvet_1998}.
This accretion-driven high-energy emission represents the main
ionizing agent at the surface of the protoplanetary disk, where the material is
transported inwards \cite{Hartmann_2016}.
There is general consensus that the main features of this picture are correct, 
but key processes remain uncertain---especially when it comes to attributing FUV 
emission to individual processes.

At FUV wavelengths, the main diagnostics of accretion are enhanced fluxes in  
hot ion lines like C~{\sc iv} (Fig.~\ref{fig:BP_Tau_Civ}). Three properties demonstrate that accretion indeed
is the root cause of the excess in these lines: 
\begin{enumerate}
    \item Monitoring campaigns 
        of individual objects such as BP~Tau show that the FUV fluxes vary in 
        conjunction with optical excess emission \cite{Simon_1990,GdC_1996,GdC_1997}.
    \item Line fluxes exceed those of WTTSs and 
        other magnetically active stars \cite{JK_2000_IUE}; still, the luminosity in, e.g., 
        C~{\sc iv} is only a tiny fraction ($\sim0.1$\%) of the 
        accretion luminosity \cite{Calvet_1996}.
    \item The lines are 
        broader in CTTSs than in WTTSs \cite{Valenti_1993}.
\end{enumerate}

\begin{figure}[t!]
\begin{center}
\includegraphics[width=14 cm]{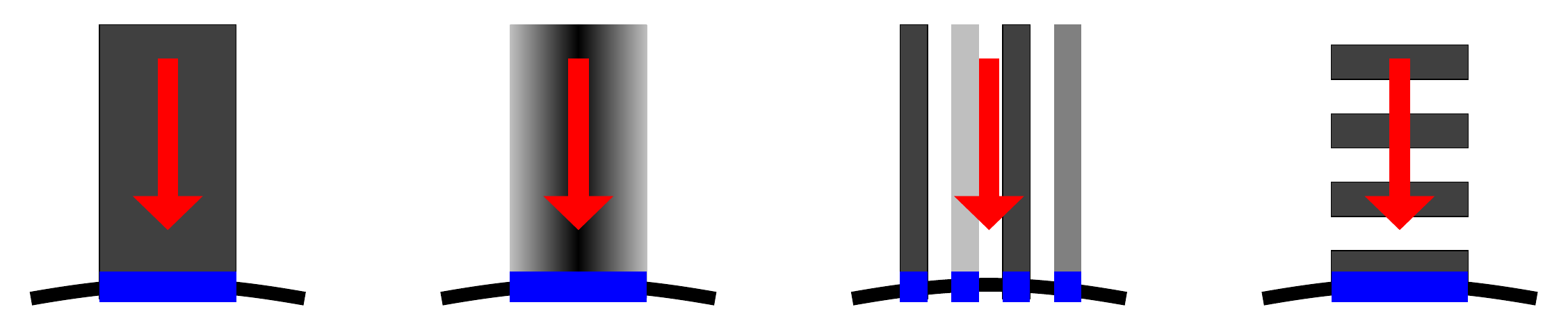}
\end{center}
\caption{Some scenarios for the 
structure of accretion columns. The stellar surface is at the bottom and 
material flows along a magnetic field line onto the stellar surface (indicated 
by the red arrow). The accretion shock (blue) forms on the stellar surface. 
Density is shown as gray scale.
From left to right: (1) One homogeneous
column with one density and a single infall velocity; (2) one column
with a density stratification (high density in the center of the column indicated
by the black region, lower density in outer region indicated by the grayish color) and one infall velocity; 
(3) multiple columns that are individually homogeneous, i.e., have different 
densities but are otherwise equal; (4) one ``column'' is decomposed into 
individual blobs that are, individually, homogeneous in density.
\label{fig:acc_col}}
\end{figure}

In particular \citet{JK_2000_IUE}  find 
that the ``emission line luminosity in the 
high ionization lines present \dots correlates well with the mass accretion 
rate.''. However, these authors also noted that the existence of the correlation 
depends critically on the source of their extinction values. 
A significant correlation results only for a certain set 
of extinction values and this strong dependence on the often uncertain UV extinction
is still relevant today. 

\begin{figure}[t!]
\begin{center}
\includegraphics[width=0.98\textwidth]{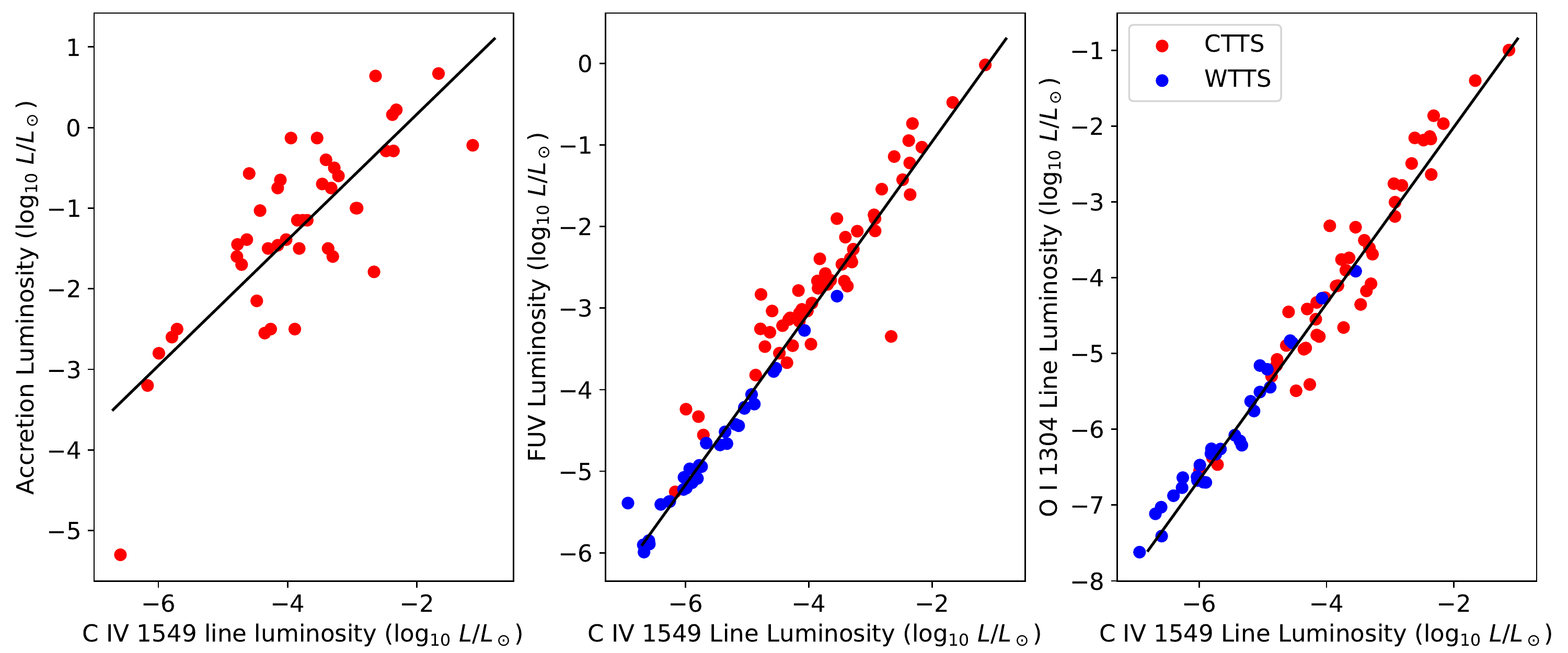}
\end{center}
\caption{{\bf Left}: Correlation between extinction corrected C~{\sc iv} and 
accretion luminosity. This suggests that the C~{\sc iv} in CTTSs is mostly 
due to the accretion process. {\bf Middle}: Correlation between C~{\sc iv} and 
total UV luminosity. {\bf Right}: Relation between line 
fluxes. Data and fits from \citet{Yang_2012}.
\label{fig:Yang_correl}}
\end{figure}

Later, the relation between 
FUV fluxes in hot ion lines and accretion was confirmed  
by \citet[see Fig.~\ref{fig:Yang_correl},][]{Yang_2012} with a large set of 91 stars with data 
from HST ACS, STIS, and GHRS lending further support to the notion that  
accretion is indeed responsible for the bright emission in hot ion lines. Interestingly, the ratio between 
fractional line luminosities of different ions, e.g., C~{\sc iv} 
and O~{\sc i}~$\lambda1304$, is the 
same in CTTSs  as in WTTSs although the respective peak line formation 
temperatures are very different ($\log~T\sim4$ vs 5) and 
although CTTS line fluxes are one to two orders of magnitudes higher
than those of comparable WTTSs 
\citep[cf. Fig.~\ref{fig:Yang_correl} right and][]{Yang_2012}.

Pinpointing the origin of the hot ion lines is challenging and currently 
an open issue; different scenarios for the accretion column have been 
proposed (cf. Fig.~\ref{fig:acc_col}). 
Most of the accretion funnel has only
$T\sim10,000$\,K \cite{Muzerolle_1998,Kwan_2011}, therefore only two regions may generate the accretion contribution to 
the hot ion emission. The material 
of the funnel immediately above the shock (the radiative precursor) or 
postshock material are the only likely sources of the hot ion line excess \citep{Ardila_2013}.

The relative importance of pre- and post-shock regions
depends mainly on shock velocity and density \cite{Lamzin_1998}. 
Both components should have red-shifted velocities:
Material in the 
funnel should have velocities of a few 100\,km\,s$^{-1}$ while the shocked
material will be at velocities of just tens of km\,s$^{-1}$ and the 
expected line profile is rather narrow (e.g.,
for the typical $v_{in}=300\,$km\,s$^{-1}$, $v_{post}=70$km\,s$^{-1}$).
However, in the Goddard High
Resolution Spectrograph (GHRS) spectra ($R\sim20,000$) of eight CTTS  
presented by \citet{Ardila_2002}, one  
finds broad emission lines 
from hot ions (C~{\sc iv}, Si~{\sc iv}, etc.) with a range of centroid
velocities (from blue-shifted to centrally peaked to red-shifted). Also higher 
excitation lines like O~{\sc vi} show a great variety of line profiles, 
again including peaks that are blue- and red-shifted from the stellar rest velocity
\cite{Guenther_2008}. 
Later 
studies using STIS \cite{Herczeg_2005} as well as 
COS data \cite{Ardila_2013} confirm these trends revealing sometimes 
complex line profiles that may be modified through absorption or emission 
by outflowing material. When the Si~{\sc iv} line profile can be measured
under the H$_2$ emission, the 
C~{\sc iv} line shapes agree relatively well. However, the line ratios between 
the blue and red members of the doublets agree with the 
ratio expected from optical thin emission
only (generally) for C~{\sc iv} and not for Si~{\sc iv}, which is puzzling given
that the radiative precursor should be optically thin to both, Si~{\sc iv}
and C~{\sc iv} \cite{Lamzin_1998} (although this implicitly assumes that 
the lines are predominately collisionally excited). Furthermore, 
the C~{\sc iv} line shape agrees with that of Si~{\sc iii}], which
is thought to come from the radiative precursor \cite{Lamzin_2000}.

In a comprehensive analysis of FUV lines from CTTSs using HST COS data, 
\citet{Ardila_2013} demonstrate that most 
hot ion lines, in particular Si~{\sc iv}, C~{\sc iv}, and N~{\sc v}, share
one kinematic profile and can be decomposed into a narrow and broad 
component. The ratio between these two components depends on the accretion
with the narrow component's contribution increasing with accretion rate.

It is interesting to note that the conversion from FUV line fluxes 
to stellar accretion rate is based on empirical correlations between
the FUV line fluxes and other, perhaps more traditional, accretion
tracers like H$\alpha$ equivalent width (EW) or veiling measurements 
\citep{JK_2000_IUE}. Application of the \citet{Calvet_1998} 
shock models by \citet{Ingleby_2013}
to the NUV to optical spectrum of CTTSs revealed that the same accretion
column (specified by their energy flux) cannot simultaneously 
reproduce the UV and optical ranges of the spectrum; these authors
require the existence of multiple accretion columns with different F-values,
i.e., different densities for the same infall velocity. With this prescription,
the emission of the different funnels peaks at different wavelengths and 
a more precise (but not necessarily accurate) description of the excess flux
is achieved, which also modifies the derived accretion rate by roughly a factor of two 
(\cite{Ingleby_2013}, \cite{Robinson_2019}). Such a scenario (multiple accretion columns with different 
densities) is appealing, because it may explain 
(a) that UV line ratios indicate rather low densities in the emitting region 
($10^{10}\,$cm$^{-3}$) based on ratios between the He~{\sc ii} and C~{\sc iv} 
lines \cite{Ardila_2013} as well as semi-forbidden lines that should be 
optically thin \cite{GdC_1999}. In addition, (b), this could explain 
the varying ratio between the narrow and broad components, because part of the 
post-shock material may be buried in the photosphere and a varying ratio 
of the pre- and post-shock plasma is seen depending on the exact accretion 
geometry \cite{Ardila_2013}. A detailed description of the accretion process 
is beyond the scope of this review, however, we think that there are not necessarily 
different accretion funnels, but rather one column with a density 
stratification (see sketch in Fig.~\ref{fig:acc_col}) and we note that structured accretion streams may also 
explain some of the X-ray diagnostics
\citep[e.g.][]{Matsakos_2013, Colombo_2016,Schneider_2018}. In this picture, the individual funnels with 
different densities are replaced by a single column, but with a continuum of 
densities so that the description of the excess emission as a sum of
different density columns is just a simplification of the true column 
structure and the accretion column may also possess
some structure along the flow direction. In summary, our picture of the accretion 
funnel evolved over the last two decades from a single monolithic column
to a structure that is stratified in density along the radius with an 
additional density modulation along the infall direction. As of 2020, however,
there is no quantitative model that simultaneously reproduces the optical/NUV/FUV 
features of accretion emission (although individual aspects can be reproduced in simulations).  
We need both simultaneous observations from the X-ray to the optical range 
as well as theoretical work to self-consistently describe the accretion process
of CTTSs.

\section{Protoplanetary Disks}
Circumstellar accretion disks are essential 
components of CTTS systems. They are typically studied
at infrared to mm-wavelengths as those are the wavelengths where most
of the energy received from the central star is re-radiated 
\cite{Williams_2011}. These so-called protoplanetary disks are not only the 
sites of ongoing planet formation, but are also highly structured and dynamic 
objects with features like warps, (dust) gaps, and spirals 
(e.g. \cite{Andrews_2018}). It is currently debated if these features can be ordered into a sequence 
in time; in particular, if so-called transitional
disks (TDs), which harbor significant (dust) gaps, are more evolved disks
compared to full gas and dust disks.

Protoplanetary disks are thought to consist
mainly of gas with just about one percent of the mass in dust like the ISM.  
The local physical properties like temperature and density 
depend, to first order, mainly on radius and height within the 
disk (see sketch in Fig.~\ref{fig:sketch}). One would like to
study both the disk gas and dust since gas controls its dynamics while the dust
is highly relevant for the formation of planetary cores. Observationally,
the dust is more readily accessible compared to the gas, because of its
continuum emission at sub-mm wavelength while observing protoplanetary disk gas requires 
more challenging spectroscopic line measurements.

The dominant species of protoplanetary disks,
molecular hydrogen, is difficult to observe directly, because the electric 
dipole transitions between rotational and vibrational states are forbidden as H$_2$ has no net dipole moment.
H$_2$ can undergo quadrupole rotational transitions that produce weak
emission lines in the mid-IR due to low oscillator strengths. 
The large spacing between even the lowest energy levels makes it, 
however, difficult to excite the molecules via collisions in cold gas of the 
disk midplane. Therefore, the IR-H$_2$ lines are weak and most
sample surveys find a relatively low detection rate with accordingly tight 
constraints on the amount of warm 
($\gtrsim 500\,$K) molecular hydrogen in the disks \cite{Bary_2008,Bitner_2008}.
Carbon monoxide (CO) is often used as a tracer for the disk's gas content, especially 
with the advent of ALMA. However, the interpretation of the CO
emission is 
challenging since CO may freeze-out on dust grains and detailed models are 
required to convert the measured emission line fluxes 
into gas disk masses \cite{Miotello_2014}. Therefore, 
complementary diagnostics of the disk's gas are highly desired.

On first sight, it may appear surprising that 
gas disk diagnostics reside in the FUV because equilibrium temperatures in the disk 
are very low (tens to hundreds of K) compared to the peak 
formation temperatures of prominent atomic emission lines (like C~{\sc iv} at 
$\sim10^5\,$K). Nevertheless, FUV observations of 
(i) fluorscently excited molecular emission lines, (ii) the 1600\,\AA{} bump,
and (iii) disk absorption features provide unique information on  
disk chemistry and dispersal.

\subsection{Fluorescently Excited Molecular Emission Lines \label{sect:H2disk}}
In contrast to the weak  IR-H$_2$ lines, the FUV transitions
of H$_2$ have large oscillator strengths. These lines result from  
electronic Lyman- and Werner-band transitions, which break symmetries in the molecular structure and enable dipole-allowed transitions. The UV-H$_2$ features originate 
from a population of vibrationally excited gas ``pumped" from 
the ground electronic state 
into the low-lying excited electronic states
by Ly$\alpha$ photons, the strongest stellar emission 
line in the FUV \cite{Schindhelm_2012}. Depending on the pumping wavelength,
molecular hydrogen is excited into the Lyman ($2p\sigma B ^1\Sigma^+_u$, 
\cite{Abgrall_1993a}) or Werner ($2p\pi C ^1\Pi_u$, \cite{Abgrall_1993b}) 
electronic bands. Each excited state emits a characteristic spectrum, called 
progression, with known ratios between the individual lines for each 
progression. The dipole-allowed transitions have large Einstein coefficients
($A_{ul}\sim10^8\,$s$^{-1}$) so that these 
electronic states will decay instantaneously in a
fluorescent cascade down to one of many different rovibrational
levels in the ground electronic state ($X^1\Sigma^+_u$, \cite{Herczeg_2002}).

\begin{figure}[t!]
\begin{center}
\includegraphics[width=0.4\textwidth]{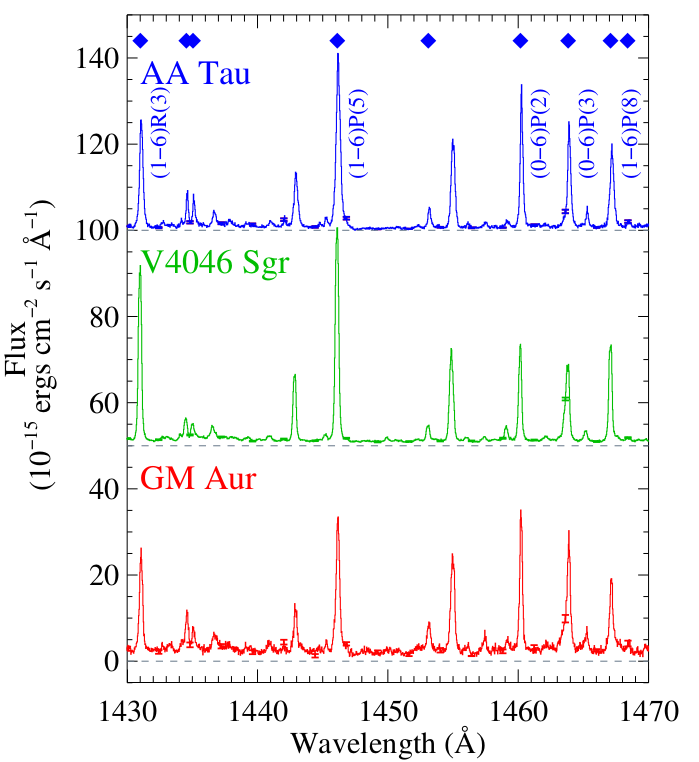}
\includegraphics[width=0.58\textwidth]{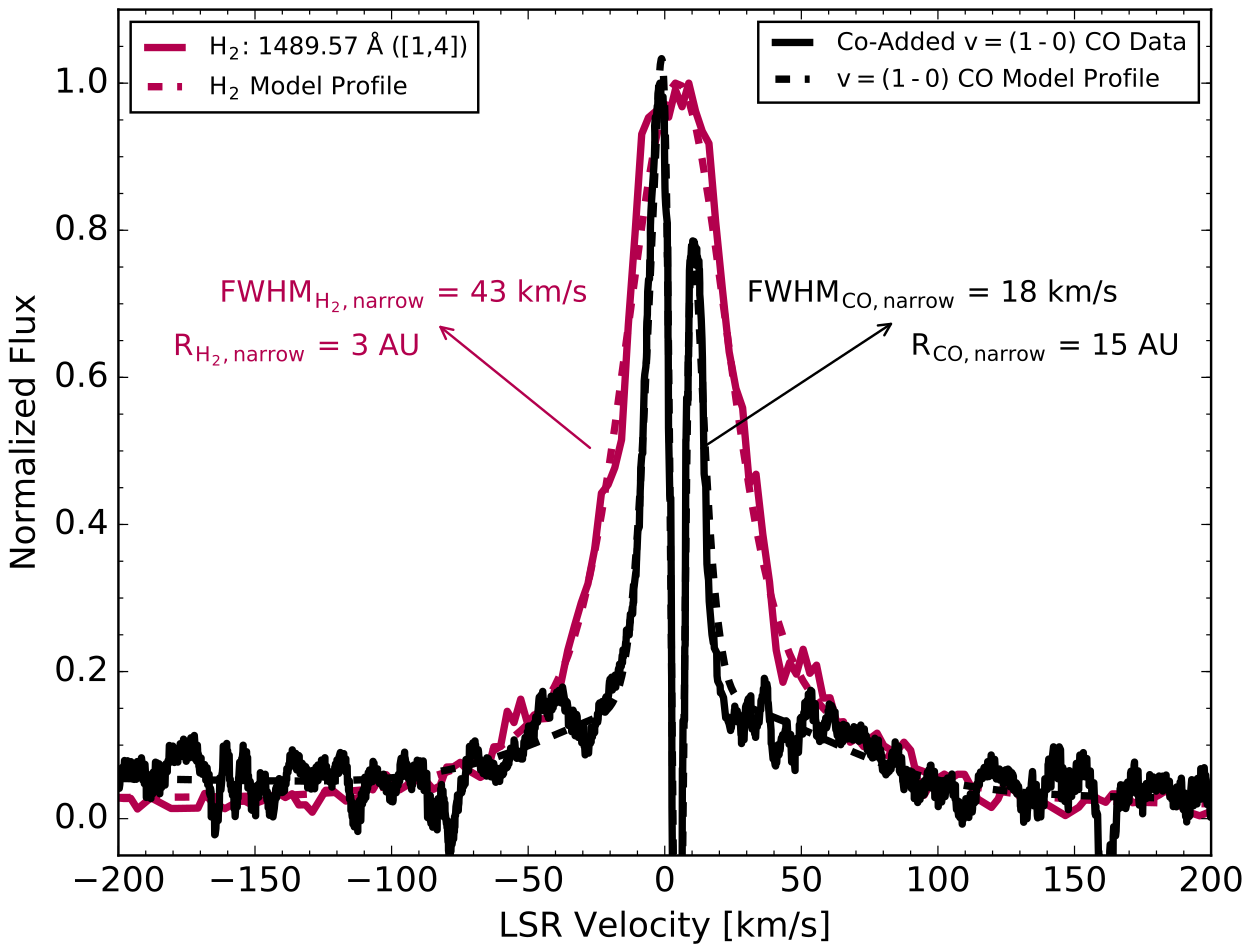}
\end{center}
\caption{{\bf Left}:  1430--1470\,\AA{} spectral region for three prototypical 
gas-rich disks. All of the strong spectral features in this bandpass are 
emission lines from Ly$\alpha$-pumped fluorescent H$_2$ (UV-H$_2$; marked with
blue diamonds and several bright features are labeled). From \citet{France_2012}.
{\bf Right}: Kinematic model for the H$_2$ emission including the effect of 
the COS line spread function (LSF). From \citet{Arulanantham_2018}. 
Both figures: \textcopyright{}AAS. Reproduced with permission.
\label{fig:H2_spec}}
\end{figure}

In protoplanetary disks, the UV-H$_2$ emission originates in the inner 
regions ($0.1 < \sim r \lesssim 10$\,au) \cite{Herczeg_2002, Herczeg_2004, France_2012, Hoadley_2015}
where gas temperatures can reach the 1500 K threshold required for Ly$\alpha$ 
fluorescence to take place \cite{Adamkovics_2016}.  A large number of
fluorescent UV-H$_2$ emission lines are observed in the 1050--1700\AA{} 
wavelength range 
\cite{Herczeg_2002,France_2012} accessible by HST observations (see 
Fig.~\ref{fig:H2_spec}). The UV-H$_2$ fluxes correlate 
with hot ion line fluxes like C~{\sc iv}, which suggests that one mechanism
controls the FUV spectrum \citep{JK_2000_IUE}---and accretion appears as the 
most probable candidate, because higher accretion rates imply more Ly$\alpha$
with a resulting higher excitation rate of the UV-H$_2$ lines.
The existence of UV-H$_2$ as well as CO strongly suggest
that both the CO and H$_2$ in the inner disk are shielded by very little neutral hydrogen 
\citep{Schindhelm_2012,France_2012b}.

\citet{Herczeg_2002} 
measured 146 UV-H$_2$ emission lines from TW Hya with \emph{HST}-STIS and 
find that the features are coincident with the star in velocity space and 
not spatially extended beyond the 0.05" resolution of the instrument, 
as would be expected for emission from an outflow. These observations indicate 
that the emitting H$_2$ is located in the inner regions of the protoplanetary 
disk (within $r \sim1.4$ AU at the distance of 56 pc to TW Hya). Nevertheless,
some contribution from an outflow may be present as 
the H$_2$ line centroids are generally slightly blue-shifted \cite{Herczeg_2005} 
and do not typically show the double-peaked profiles expected from a pure 
disk emission model \cite{Hoadley_2015}, see also sect.~\ref{sect:jets}.

\begin{figure}[t!]
\begin{center}
\includegraphics[width=0.5\textwidth]{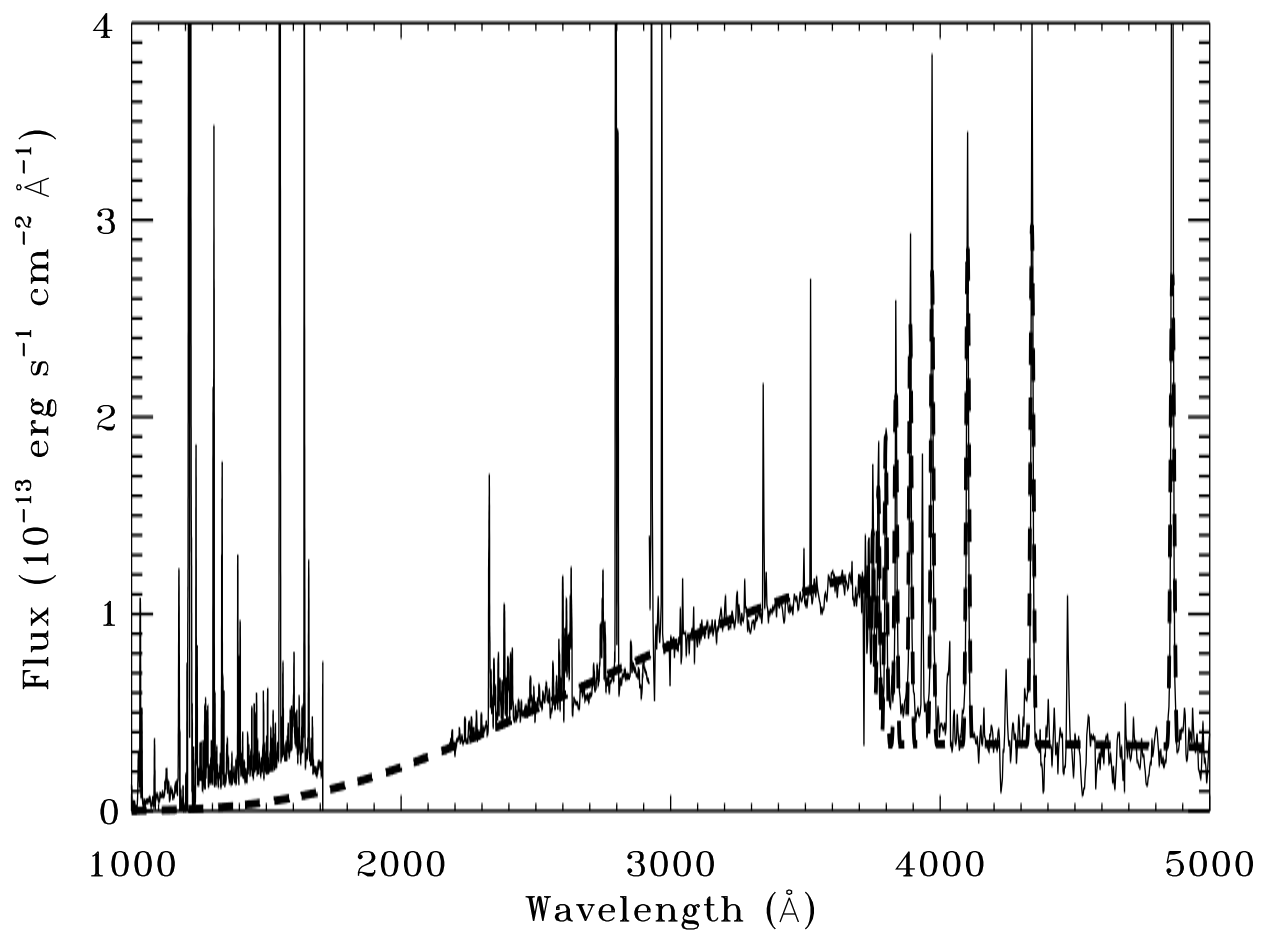}
\includegraphics[width=0.49\textwidth]{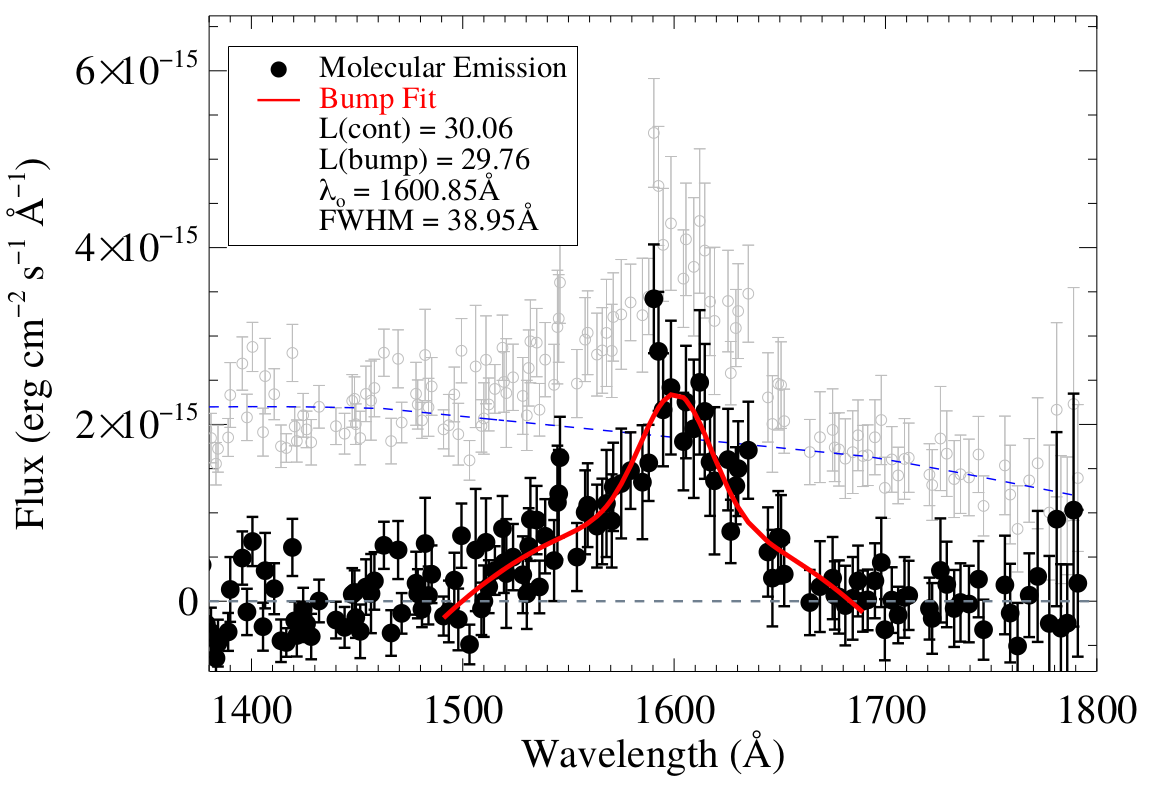}
\end{center}
\caption{{\bf Left}: FUV to optical spectrum of the nearest CTTS, TW~Hya, 
after subtraction of WTTS template spectrum of V819~Tau. The dashed line 
indicates the expected emission from the accretion model, which approximates the
NUV continuum of TW Hya but significantly underestimates the FUV continuum, 
especially around 1600\,\AA{}.
From \citet{Herczeg_2004}.
{\bf Right}: Typical 1600\,\AA{}-Bump spectrum, here for CS~Cha. The blue-dashed 
line indicates the polynomial model used to subtract the accretion continuum.
From 
\citet{France_2017}. Both figures: \textcopyright{}AAS. Reproduced with permission.
\label{fig:H2_bump}}
\end{figure}

In CTTSs, one observes a number of H$_2$ absorption features against the 
wings of the stellar Ly$\alpha$ line. The relative depths of 
these H$_2$ absorption features depend on the respective absorbing
column density $N(H_2, \nu, J)$. The relative population of these states
depends on the total column density and the excitation temperature (assuming a single, thermal population of absorbing molecular hydrogen \cite{France_2012c}).
Measuring numerous absorption features for 22 CTTSs, \citet{Hoadley_2017}
find that (a) the low energy states follow the pattern expected for 
$T\sim2,000$\,K while the high-energy states of H$_2$ show higher 
occupancies than expected.  These authors speculate that the non-thermal populations arise in a diffuse molecular disk atmosphere where level populations are driven out 
of thermal equilibrium by Ly$\alpha$ photon pumping.

Kinematic modelling is another powerful technique to
derive spatial information from the UV-H$_2$ emission lines as the
spectral resolution of the UV instruments onboard HST is sufficient 
given the velocities expected in the inner disk region
 \cite{Schneider_2015,Hoadley_2015}. 
Assuming that the emission at each radius has a Keplerian profile, the radial 
profile of the molecular hydrogen emission can be reconstructed, see 
Fig.~\ref{fig:H2_spec}. The emission is confined to 
$r<10\,$au for most sources with a characteristic emission radius of just 
about 1\,au, which agrees with 
the upper limits imposed by the spatial analysis of STIS data 
\cite{Herczeg_2006}. The UV-H$_2$ emission can be also used to study 
the evolution of disk gas with system age as described in 
more detail in sect.~\ref{sect:dissipation}.

The second most abundant species in protoplanetary disks, CO, also emits in the FUV. Photo-excited CO gas traces cooler 
disk regions  compared to H$_2$ (UV-CO; $T\sim200-500$\,K). Its emission
originates from several ro-vibrational bands of the $A^1\Pi - X^1 \Sigma^+$ 
(Fourth Positive) electronic transition system. Similar to the UV-H$_2$, 
the UV-CO emission is pumped by stellar (accretion driven) Ly$\alpha$ and 
partly by C~{\sc iv} photons. \citet{Schindhelm_2012} detect UV-CO emission
from $T\sim500\,$K
gas in roughly half of the FUV disk spectra, i.e., 
a different population than the warmer CO gas ($T\sim300-1500$\,K) well studied 
in the strong fundamental band rovibrational emission lines 
at 4.7-5 $\mu$m \cite{Salyk_2009,Salyk_2011,Brown_2013}. These differences may be attributed to the FUV 
CO population residing more distant from the star than either the 
IR-CO or UV-H$_2$ emitting gas \citep{Schindhelm_2012}. 
Compared to UV-H$_2$, the signal 
of the UV-CO lines is weaker and kinematic modelling subject to larger 
uncertainties. Generally, 
however, the UV-CO emission correlates with the stellar Ly$\alpha$ emission, 
i.e., mainly with accretion rate. The second strongest parameter controlling the 
the UV-CO emission may be the dust and gas opacity of the inner disk, or in
other words, observations of UV-CO emission is more indicative of the dispersal of cooler gas than UV-H$_2$ emission.

\begin{figure}[t!]
\begin{center}
\includegraphics[width=0.39\textwidth]{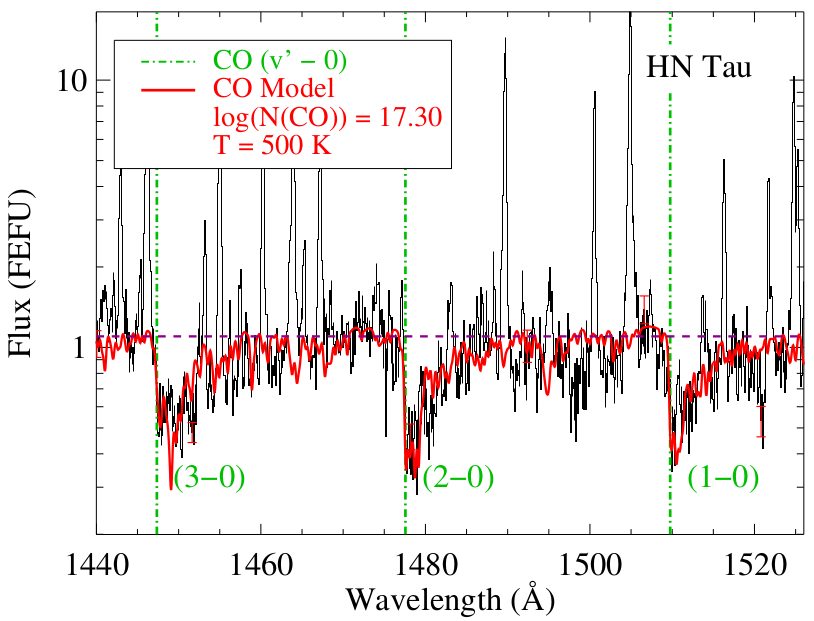}
\end{center}
\caption{ CO absorption in the CTTS HN~Tau.
Data shown in black, fit of the CO transmission spectrum in red and the green 
line indicates the bandheads (1 FEFU$=10^{-15}$\,ergs\,cm$^{-2}$\,s$^{-1}$
\,\AA$^{-1}$). From 
\citet{France_2011}. \textcopyright{}AAS. Reproduced with permission.
\label{fig:CO_emiss}}
\end{figure}

\subsection{The ''1600\,\AA{} Bump'' \label{sect:bump}}
The FUV continuum emission of CTTSs is comparably weak, because both, the 
stellar and the accretion continuum, decline towards shorter
wavelengths; in  contrast, the accretion continuum is readily detected in the NUV 
\citep{Ingleby_2013}. Therefore, good continuum measurements  require high S/N
and the discovery of a broad (50--200\,\AA{}) emission feature, 
roughly centered on 1600\,\AA{} in the FUV continuum of CTTSs
was somewhat surprising 
\citep[see Fig.~\ref{fig:H2_bump} and refs ][]{Bergin_2003,Herczeg_2004,Ingleby_2009}. In fact, 
the dominant bump formation is still not entirely clear, but most likely results 
from line and continuum emission from vibrationally excited H$_2$.

\citet{France_2017} detect the 1600\,\AA{}-Bump in 
19 out of 37 CTTSs surveyed. The detection rate is 100\,\% in transition disks 
compared to about one third in primordial or non-transition disk sources.
Because the integrated flux of the 1600\,\AA{}-Bump is a significant fraction
of the entire H$_2$ and continuum flux, \citet{France_2017} suggest that the
bump draws its energy from the Ly$\alpha$ radiation field. Specifically, 
these authors suggest that the Ly$\alpha$-driven dissociation of water in 
the inner disk generates a population of highly non-thermal H$_2$ and emission from this 
population then produces the 1600\,\AA{}-Bump, i.e., efficient
water dissociation by Ly$\alpha$ photons during the later stages of disk
dispersal \cite{France_2017}. This scenario is supported by the lack of a 
correlation between 1600\,\AA{}-Bump and stellar X-ray luminosity 
\cite{Espaillat_2019}.
Another cause for large 1600\,\AA{}-Bump luminosities may be high photo-electron
densities (presumably driven by stellar X-rays; \cite{Bergin_2003, 
Ingleby_2009}).

\subsection{Disk Absorption}
For moderately inclined disks ($i\gtrsim70^\circ$), the line of sight towards
the central star passes through the upper disk regions. The 
observed spectrum then contains absorption features from the disk. In the FUV,
the two most abundant species in protoplanetary disks, H$_2$ and CO, have 
strong absorption bands, which allow the measurements of column densities and
rovibrational temperatures. The absorption profile depends on (1) an (assumed) 
turbulent velocity (typically small, $\sim0.1\,$km$^{-1}$), (2) 
a column density, and (3) a rotational temperature. 

While FUV-CO absorption is an established technique in many fields of astronomy,
e.g., for the study of the interstellar medium (ISM, see \cite{Burgh_2007} and
references therein) or planets \cite{Feldman_2000}, its use for protoplanetary 
disks was first proven in \citeyear{France_2011} by \citet{France_2011}, who
measure absorption by the ground vibrational state ($\nu$--0) of CO against 
the backlight of the FUV accretion continuum of the central star. 
In their analysis of 34 T~Tauri stars, \citet{McJunkin_2013} find CO absorption in 
about one quarter of the sources. For these sources, the 
FUV-CO absorption velocities are compatible with the radial velocity of the 
target star within the instrumental velocity calibration uncertainty of 
about 15-20\,km\,s$^{-1}$.
The data imply rotational temperatures around 500\,K, which may be reached out 
to radii of 0.2--2\,au assuming thermal equilibrium with the stellar 
radiation field. Compared to the IR-CO emission, the temperatures of the UV-CO 
absorption are lower, perhaps because FUV-CO absorption traces a gas population 
at larger disk radii than the IR data. When combined with estimates for the
conditions required for a thermal equilibrium between the rotational states
and the radiation field from disk models \cite{Woitke_2011}, 
\citet{McJunkin_2013} suggest that the CO absorbing gas resides in disk layer
at a height of $z/r\sim0.6$ and 0.7, i.e., the flared upper disk atmosphere.

In the FUV spectra of CTTSs, H$_2$ absorption is seen both, against the 
continuum and against the Ly$\alpha$ line. For estimating an CO/H$_2$-ratio, 
however, one should compare the respective absorption features seen against the 
continuum \citep[see discussion in][]{France_2014}. In the best characterized CO and H$_{2}$ absorption system, RW Aur, the CO/H$_2$ ratio is found to be $\approx 10^{-4}$, which is compatible with the value 
for the ISM. Since the CO/H$_2$ is of fundamental importance
for disk chemistry and disk mass estimates, future studies using FUV absorption 
will provide unique information not attainable with, e.g., 
ALMA observations and disk modeling. Therefore, future mission concepts like
NASA's LUVOIR specifically aim at  high-sensitivity, high-resolution absorption 
line spectroscopy through moderately inclined disks \citep{France_SPIE}.

\subsection{Disk Chemistry and the UV SED}
The strength and shape of the FUV radiation field has a strong
influence on the chemical abundances of the disk, both at 
planet-forming radii \citep[$r < 10$\,au;][]{Bergin_2003, Walsh_2012} and at larger 
radii ($r>50\,$au) where the majority of the disk mass resides 
\cite{Bergin_2007}. The stellar FUV continuum controls
the dissociation of the most abundant disk molecules (H$_2$ and CO; 
\cite{Shull_1982,Dishoeck_1988}). The propagation of the FUV continuum is mainly 
regulated by dust grains \cite{Zadelhoff_2003}; the processes of grain-growth 
and settling likely allow these photons to penetrate deeper into the disk as 
the protoplanetary environment evolves \cite{Vasyunin_2011}.

\begin{figure}[t!]
\begin{center}
\includegraphics[width=0.65\textwidth]{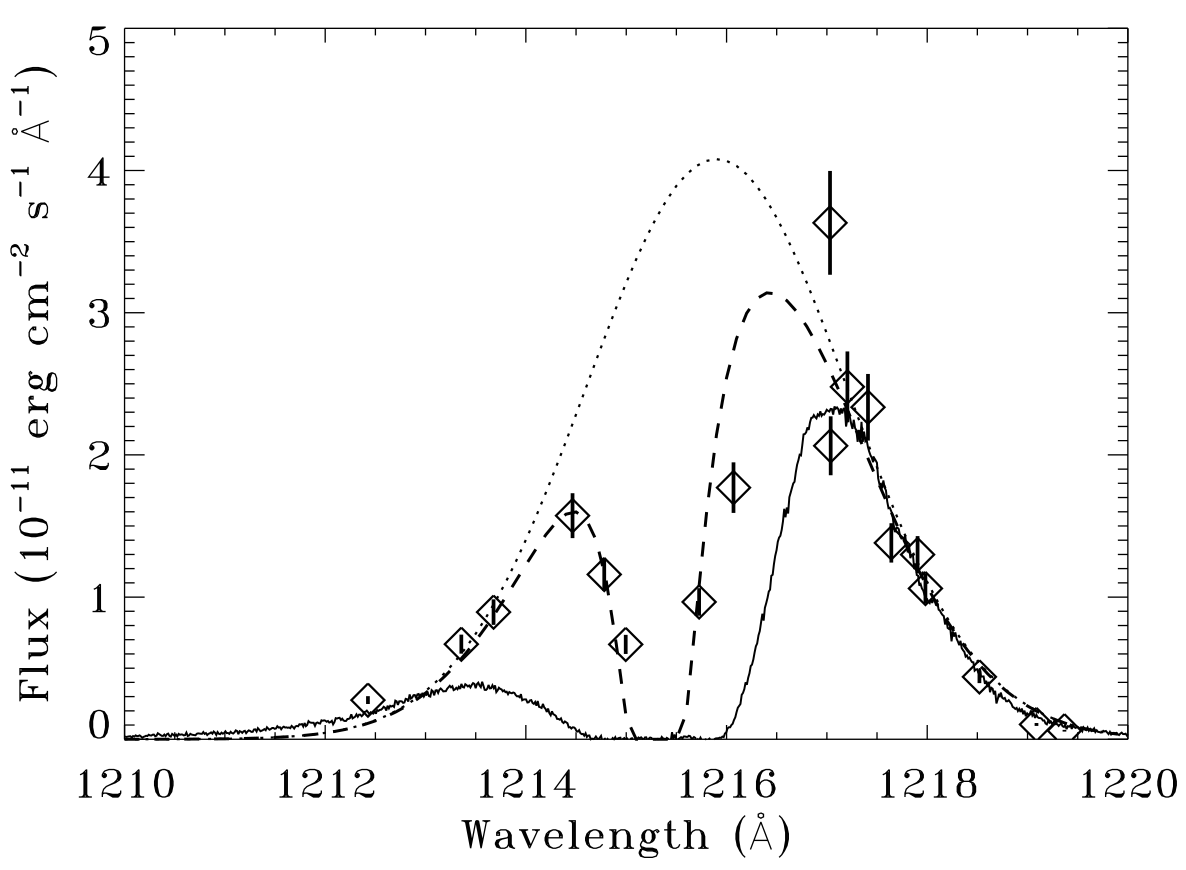}
\end{center}
\caption{Example of a Ly$\alpha$ reconstruction based on H$_2$ fluxes. Diamonds
indicate the wavelengths at which measured H$_2$ progressions are pumped, the 
dashed line indicates an H$_2$ model for T=2500\,K and $\log N(H_2)=18.5$ and 
a surface filling factor for the H$_2$ material as seen by the Ly$\alpha$ 
emitting region. In this model, the intrinsic Ly$\alpha$ line is Gaussian-shaped
(dotted), the H$_2$ gas sees the dashed profile, and the Ly$\alpha$ profile 
observed by HST is subject to additional interstellar absorption (solid line).
From \citet{Herczeg_2004}. \textcopyright{}AAS. Reproduced with permission.
\label{fig:Lya_recon}}
\end{figure}

\citet{Bergin_2003} first emphasized the importance of accretion-generated 
H~{\sc i} Ly$\alpha$ to the disk chemistry, and more recently it has been 
shown that the FUV spectral energy distribution of all CTTSs is overwhelmingly 
dominated ($\gtrsim$ 80\%) by Ly$\alpha$ emission \cite{Schindhelm_2012,
France_2014}. Unlike the FUV continuum emission, the radiative transfer of 
Ly$\alpha$ photons is controlled mainly by resonant scattering in the upper, 
atomic disk atmosphere \cite{Bethell_2011}. Subsequent detailed disk modeling
has demonstrated the importance of properly accounting for Ly$\alpha$ radiation 
from the central star, finding significant ($\gtrsim1$ order of magnitude)
depletions in the abundances of C$_{2}$H$_{4}$, CH$_{4}$, HCN, NH$_{3}$, and 
SO$_{2}$ when Ly$\alpha$ is included \cite{Fogel_2011}. Interestingly, some
species with large photo-absorption cross-sections in the region around 
Ly$\alpha$, such as  H$_{2}$O, do not show significant
depletion, because the enhanced dissociation rate is balanced by 
Ly$\alpha$-driven photodesorption
of water molecules from dust grains. It is clear now that Ly$\alpha$ is a 
mandatory component of FUV radiation fields used for chemical modeling. 
Historically, the large CTTS spectral atlases do not cover 
Ly$\alpha$ \cite{Yang_2012}, or are dominated by geocoronal 
emission, such as archival IUE data. Reconstructing the Ly$\alpha$ radiation 
field impinging the inner disk region requires measurements of a large number 
of UV-H$_2$ lines (see Fig.~\ref{fig:Lya_recon}); the largest 
such atlas of high-resolution FUV spectra of CTTSs  is provided by 
\citet{France_2014} with high-level data available for download. 

Deriving accurate FUV fluxes is challenging, because extinction estimates
are typically based on the comparison between a stellar template (augmented
by a suitable accretion spectrum for CTTSs) and observed colors or 
flux-calibrated spectra plus an assumed shape of the extinction curve.
It is known that the extinction values derived from optical and NIR 
data can differ, but this effect is mild compared to the extrapolation
into the FUV where the correction factors are large. Specifically, it 
was found that the intrinsic, de-reddended UV-H$_2$ luminosity is 
correlated with the assumed extinction values, which is not physically
expected and suggests that the large extinction corrections, or rather 
errors in them, dominate over the intrinsic variations between the 
systems \citep{France_2017}. In an attempt to remedy this situation, 
\citet{McJunkin_2016} performed detailed modelling of the UV-H$_2$ lines 
including self-absorption and use the known branching ratios to recover
the shape of the FUV extinction curve. Although the line of sight 
towards the UV-H$_2$ emitting region in the inner disk may be subject
to a different extinction than the central star, \citet{McJunkin_2016}
find a number of sources for which this and more traditional methods
agree, but they also find systems for which both methods strongly
disagree and more investigation on the subject of FUV extinction curves
relevant for CTTSs is needed.

\subsection{Disk dispersal\label{sect:dissipation}}

\begin{figure}[t!]
\begin{center}
\includegraphics[width=0.65\textwidth]{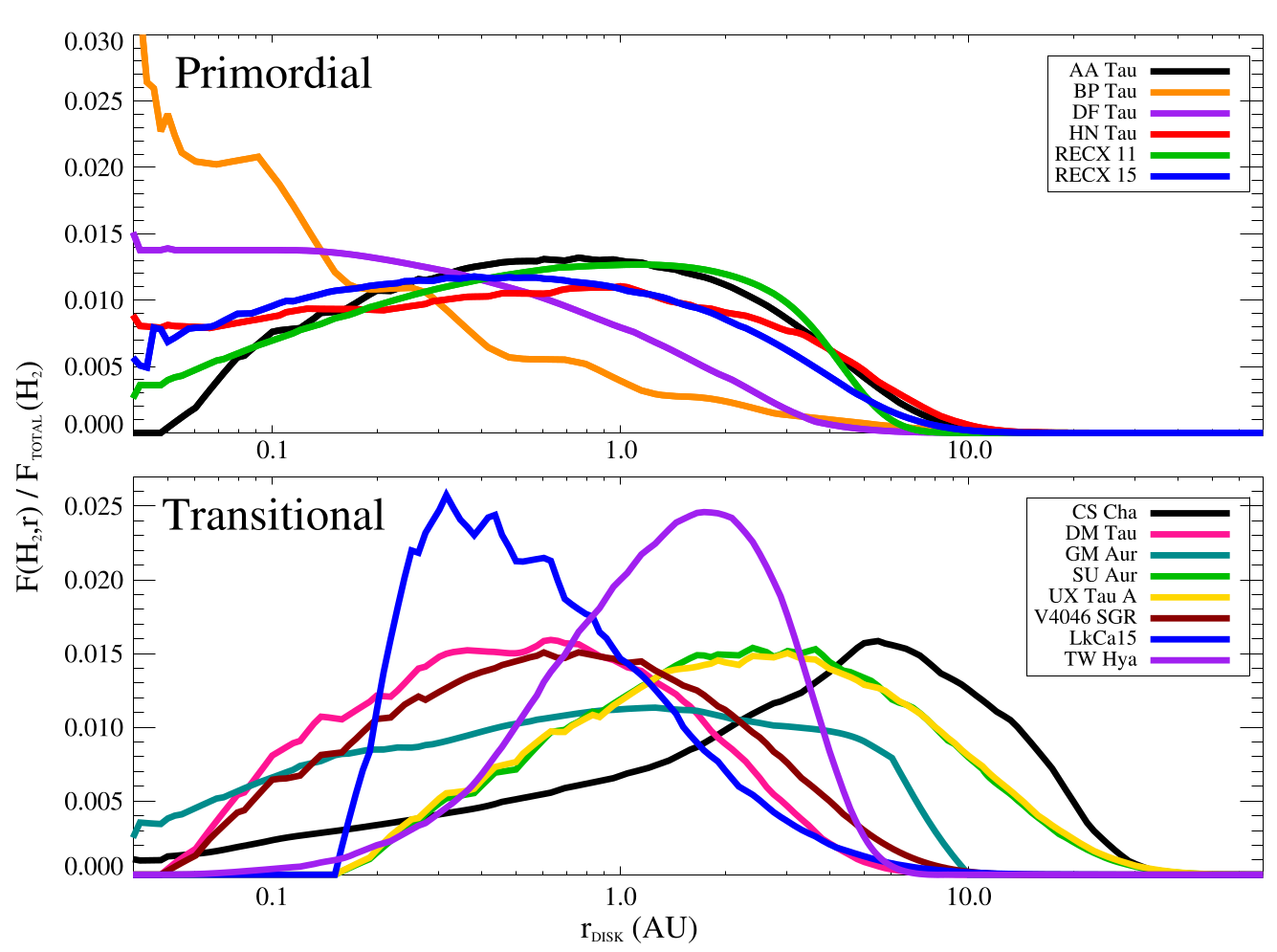}
\end{center}
\caption{Comparison of full and transitional disks in their reconstructed 
radial distributions of the observed H$_2$ emission. From \citet{Hoadley_2015}.
\textcopyright{}AAS. Reproduced with permission.
\label{fig:H2_evo}}
\end{figure}   

Primordial gas disks are known to dissipate on timescales of $\lesssim10$\,Myr,
at which point mass accretion onto the central star halts \cite{Fedele_2010}. 
At some time in the evolution of protoplanetary disks thermal winds 
carry away more mass than is transported inwards, which leads
to the dispersal of the disk from the inside-out  \cite{Alexander_2006}.
Inner gas disks may survive longer than the typical $\sim3$ Myr lifetime of
inner dust disks (e.g., \cite{Salyk_2009,France_2012b}).
The detailed physical process that clears the inner disk are not yet established. Among 
the proposed mechanisms are photoevaporation \cite{Alexander_2006,Gorti_2009} 
and dynamical clearing by exoplanetary systems \cite{Dodson_2011}, possibly 
aided by the magnetorotational instability \cite{Chiang_2007}. 

Models that simultaneously treat FUV, EUV, 
and X-ray irradiation from the central star have shown that the FUV illumination 
can control the total evaporation rate (and hence the disk lifetime) by
driving the heating at intermediate ($r\sim3-30\,$au) and large radii 
($r\geq100\,$au, \cite{Gorti_2009}). FUV radiation also
controls the gas temperature at the base of the evaporative flows through 
the generation of photoelectrons released by FUV-illuminated dust grains. 
Grain-growth and dust settling during the first stages of the planet 
formation process enable deeper penetration of the FUV radiation.

UV photons not only regulate disk dispersal, but are also powerful 
diagnostics thereof. For example, UV-H$_2$ emission is sensitive to gas surface 
densities lower than $10^{-6}$\,g \,cm$^{-2}$ \cite{France_2012}, making them an extremely useful 
probe of remnant gas during the late stages of disk dispersal. 
\citet{Hoadley_2015} study the radial distribution of the UV-H$_2$ using 
kinematic modelling of the H$_2$ profiles. In particular, these authors compare
the radial distributions of full and transitional disks (Fig.~\ref{fig:H2_evo})
finding that the distribution in transitional disks is shifted towards larger
radii. The H$_2$ and CO gas populations decline with dust disk 
dissipation \cite{Banzatti_2015,Hoadley_2017}, but the H$_2$ depletion 
lags behind the CO for disks with large inner dust cavities 
\cite{Arulanantham_2018}.

\section{Jets and outflows \label{sect:jets}}
CTTSs, and their younger siblings, the class~0 and I sources, drive outflows 
and jets (collimated ouflows, see \citet{Frank_2014} for a recent review). The 
origin of the various outflow components is not clear and different parts of the 
young stellar system may contribute with varying fractions depending on the 
actual tracer (Fig.~\ref{fig:sketch}). Specifically, the star itself
may drive a stellar wind \cite{Zanni_2013}, the region (magnetically) connecting 
the star and the disk may eject a so-called X-wind \cite{Shu_1994}, and the 
disk itself may launch a disk wind \cite{Blandford_1982}. 

Outflows and jets are traditionally studied in forbidden emission lines tracing
thin ($\log n\sim2\dots4$ in units of cm$^{-3}$) plasma with temperatures 
around $10^4\,$K and spectacular jet images exist, which show beautiful phenomena 
such as bow shocks, wiggling, and precession (e.g., \cite{Raga_2015}). The 
jet emission is usually concentrated in so-called knots where the jet material
has been shock heated and is now seen in emission lines. These knots move along 
the jet axis away from the driving source.
Cooler jet 
components are seen as molecular emission lines, which can be studied at 
radio-frequencies even for highly embedded sources \cite{Loinard_2013,Guedel_2018}. 
In recent years, the observing window for jet studies opened significantly and now
spans from the highest photon energies in the X-ray regime 
\citep{Pravdo_2001,Guedel_2008,Schneider_2011} down to m-wavelength 
radiation \citep{Ainsworth_2014}. 

Despite this large arsenal of observing opportunities, FUV observations of 
jets are highly powerful, because they allow us to measure (a) the molecular 
content through H$_2$ emission lines ($T\sim10^3$\,K) and (b) 
kinematically resolve the hot jet
in high excitation lines like C~{\sc iv} ($T\sim10^5$\,K) with very high 
angular resolution below 0.1\,arcsec \citep{Schneider_2013}. Often, there is 
indirect evidence for an outflow component in, e.g., C~{\sc iv}.
\citet{Dupree_2005} suggest that a hot ($\sim10^5$\,K) wind exists in TW~Hya
based on a roughly P~Cygni-like asymmetry observed in hot lines like 
C~{\sc iii}, C~{\sc iv}, and O~{\sc vi}. The absorption extends out to 
blue-shifted velocities of up to 400\,km\,s$^{-1}$ and \citet{Dupree_2005}
estimate that a mass-loss rate of roughly 10\,\% of the accretion rate may 
be sufficient to cause the observed absorption signals. This interpretation,
however, heavily relies on the assumption that the intrinsic line profiles 
are symmetric and that the asymmetry is indeed caused by absorption. 
\citet{JK_2007} dispute this assumption and show that other features in the line 
profiles that would be expected in a hot wind scenario are not present. They 
suggest that there is no
indication for absorption by a hot wind in TW~Hya, only for absorption in low 
ionization lines by a comparably cold wind is seen.

\begin{figure}[t!]
\includegraphics[width=0.49\textwidth]{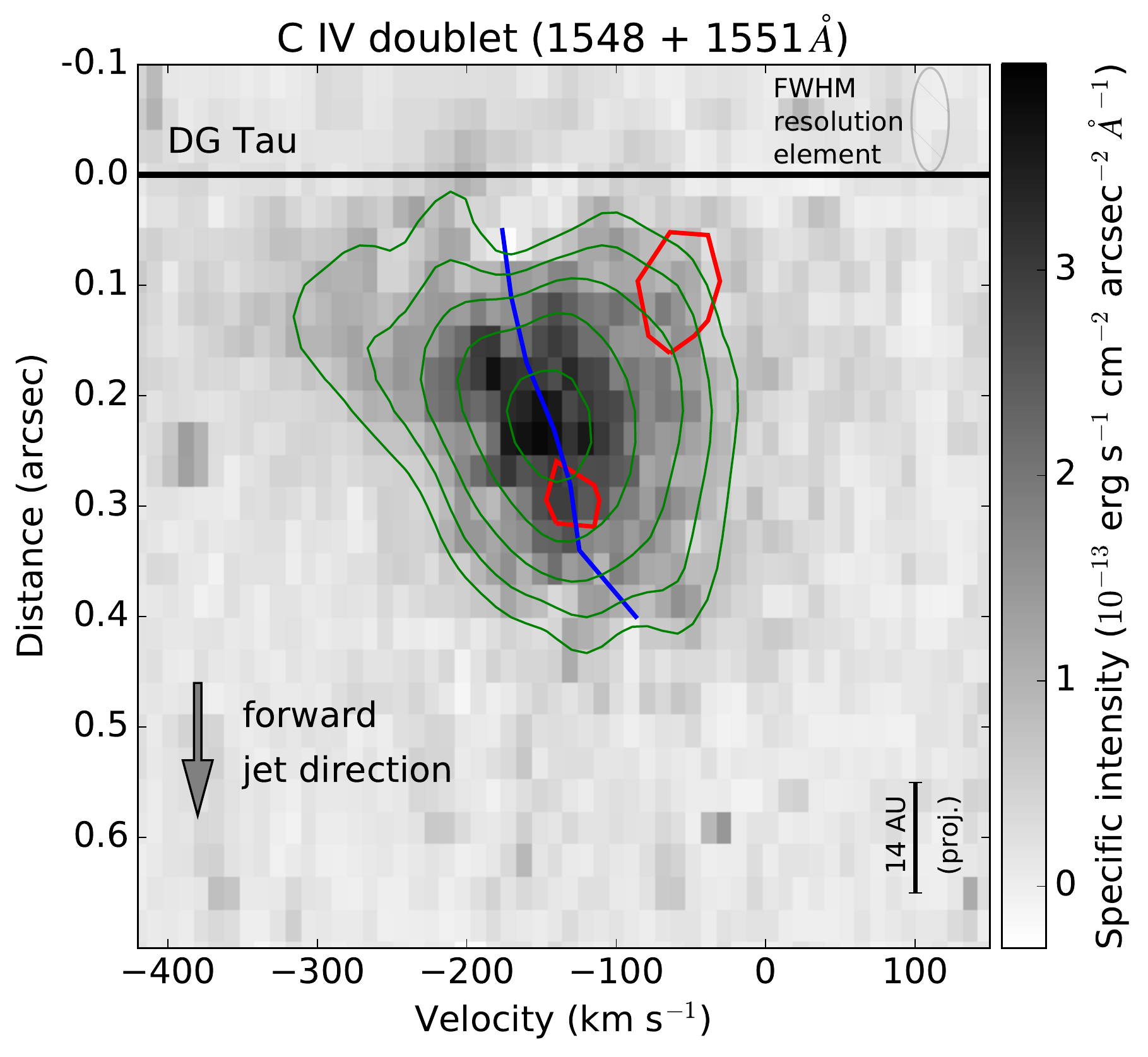}

\vspace*{-6cm}\hspace{0.5\textwidth}\includegraphics[width=0.49\textwidth]{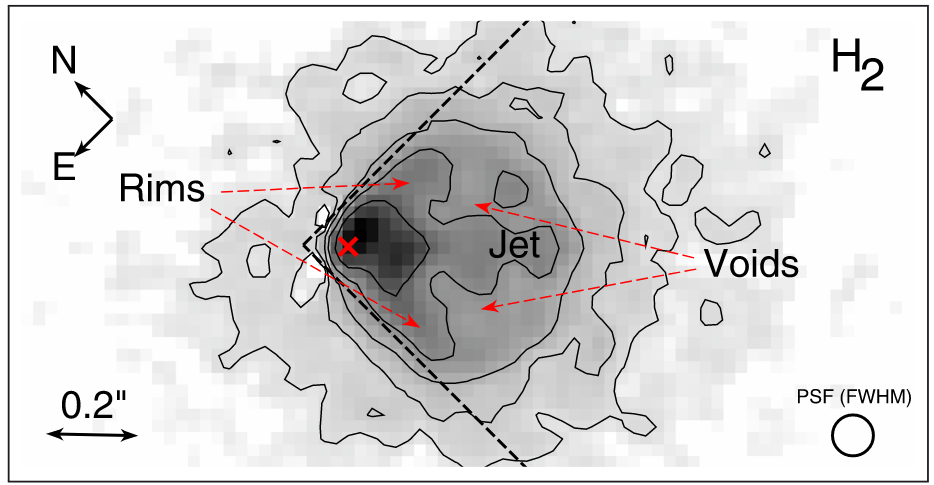}

\vspace*{2cm}
\caption{{\bf Left}: Position-Velocity diagram for the  C~{\sc iv} 
emission of the DG~Tau jet. The red contours indicate emission components 
seen in  traditional tracers like [O~{\sc i}]. Note that there is no 
emission at the stellar position and all C~{\sc iv} emission originates
within the jet. From \citet{Schneider_2013}. {\bf Right}: Image of 
UV-H$_2$ of DG~Tau. Reproduced with permission \textcopyright{} ESO.
\label{fig:jets}}
\end{figure}

While these arguments address a hot stellar wind, 
jet emission in C~{\sc iv} is directly observed and there is often also indirect 
evidence for such a hot jet, e.g., strong 
emission seen in IUE data  but not in smaller aperture HST data or the hot 
lines exhibit a similar
kinematic morphology as forbidden lines that trace the jet rather than 
accretion. Studies using long-slit data were 
initiated based on these indirect indications. 

It was found that the jet's C~{\sc iv}  emission is typically
observed to be offset from the star along the forward facing (approaching)
jet by tens of au (see Fig.~\ref{fig:jets}). Since C~{\sc iv} and X-ray emission 
spatially overlap, it is natural to assume that they probe the same jet 
component \cite{Schneider_2013}. The mass-loss associated with the C~{\sc iv} 
emission is somewhere between total mass-loss and the mass-loss related to 
the X-ray emission \cite{Skinner_2018}. 

Both, the X-ray and the C~{\sc iv} were not predicted by models and there 
is no consensus on their origin. There are, however, a number of observational
constraints that set these jet features apart from the well known optical jet.
First, the C~{\sc iv} and X-ray emission are stationary so that heating by the 
same shocks that cause the optical knots
is unlikely \citep{Schneider_2008}. Second, adiabatic cooling would have 
lowered the temperature  
of outflowing material with any reasonable initial temperature ($T<10^7\,$K)
below C~{\sc iv} emitting temperatures so that local heating is required
to power the C~{\sc iv} and X-ray emission at the measured distance 
of tens of au. Third, the velocity of the C~{\sc iv}  emission is comparable
to that of the bulk of the optical jet but spatially offset indicating a 
different heating mechanism than shocks. \citet{Schneider_2013} speculate
that magnetic heating similar to processes in the stellar coronae may be
responsible for the C~{\sc iv} and X-ray emission. In addition, a
velocity just slightly higher than the optical jet suggests that this 
outflow originates in the inner 
disk region, i.e., extends the jet's  onion-like velocity structure
to the innermost regions very close to the star where stellar and disk 
magnetic fields interact. C~{\sc iv} from protostellar jets may therefore 
be a powerful probe of the magnetic field in a accretion funnel region. 

In addition to the very hot jet, FUV observations also trace the jet's 
molecular hydrogen emission \citep{Walter_2003, Schneider_2013b}. This 
UV-H$_2$ emission is caused by the 
very same processes that cause UV-H$_2$ disk emission (cf. 
sect.~\ref{sect:H2disk}), i.e., fluorescent emission pumped by 
Ly$\alpha$ photons.  The wide-angle
structure seen in UV-H$_2$ (Fig.~\ref{fig:jets} right) appears 
stationary and agrees very well with the spatial  morphology of the H$_2$
emission seen in the NIR with AO-equipped IFUs \cite{Agra_2014}. 
The molecular outflow appears stationary in time contrasting 
the moving knots seen in optical data of protostellar jets. 
Thus, the UV-$H_2$ likely arises in a wind region
heated by ambipolar diffusion with subsequent pumping by stellar Ly$\alpha$
photons and likely 
traces probably the outermost regions of the MHD wind \cite{Schneider_2013b}.

\begin{figure}[t!]
\begin{center}
\includegraphics[width=0.8\textwidth]{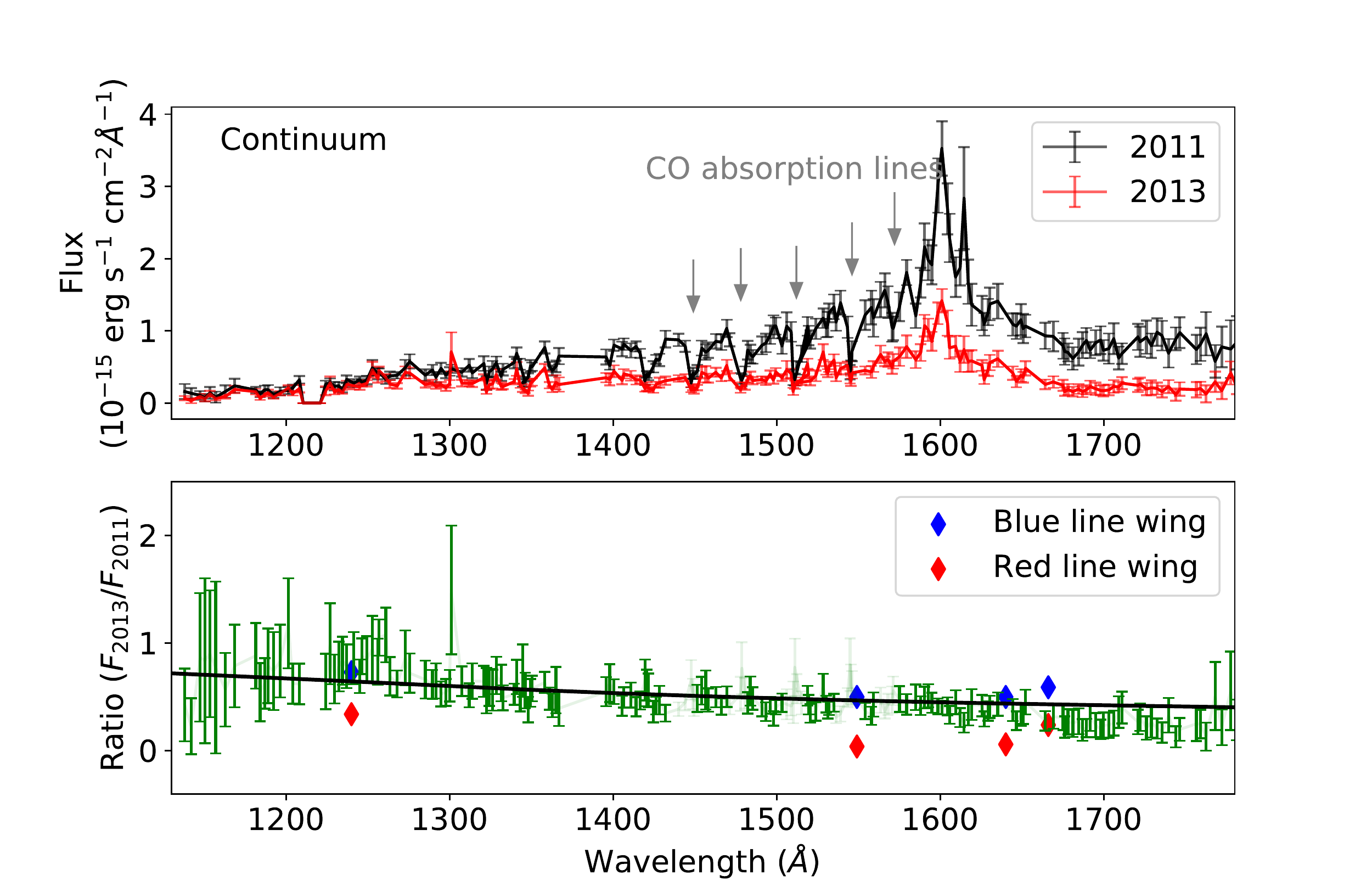}
\end{center}
\caption{{\bf Top}: FUV continuum spectra of AA~Tau during bright and dim 
states. {\bf Bottom}: Ratio between bright and dim states with the 
evolution of emission lines shown in blue/red diamonds for the 
blue and red line wing, respectively. Notable is the bluer appearance of
the continuum during the dim state, which is incompatible with just extra 
absorption and more compatible with scattering.
From \citet{Schneider_2015}. Reproduced with permission \textcopyright{} ESO.
\label{fig:AA_Tau}}
\end{figure}

\section{Variability \label{sect:variability}}
In the previous sections, we described how FUV emission can originate
from chromospheric activity, from accretion, and from outflows or jets.
All of these sources are variable, just on different time scales.
Given the general sparsity of FUV observations (since space-based
instrumentation is required) and the fact that, compared with other
wavebands, relatively low column densities of intervening material
absorb nearly all intrinsic signal, variability in the FUV band is not
well-studied.

One characteristic time scale in stars is the rotation period. Active
regions or accretion spots rotate in and out of view. This accounts partially
for observed changes in the accretion rate with time
\cite{Costigan_2014}. Early IUE monitoring of DI~Cep and BP~Tau shows that this
holds true in the UV. FUV line fluxes in DI~Cep associated with the accretion 
hot spot change over the course of a rotational cycle and additionally line and
continuum fluxes change as the accretion rate changes when comparing spectra
taken more than a decade apart. These changes are correlated with the $V$ band
magnitude \cite{GomezdeCastro_1996}. Similarly, in BP~Tau emission lines that 
are formed through recombination, i.e.\ in the accretion shock, such as 
O~{\sc i} and He~{\sc ii} vary with stellar phase, while lines that can be 
formed both in the accretion shock and in the transition region, such as the 
C~{\sc iv} doublet at 1548 and 1550~\AA{}, show less pronounced phase variation
\cite{GomezdeCastro_1997,Ismailov_2010}. In a more recent observation of
GM~Aur model fits allow to disentangle accretion rate and spot area, indicating
that, at least on this star, it is in fact the density of the accretion stream,
that changes, not the size of the shock \cite{Ingleby_2015}.

It is worth noting that the shock is also invoked as the source of soft X-ray
emission \cite{Lamzin_2004,Kastner_2002,Brickhouse_2010}, yet X-ray and FUV
flux appear uncorrelated \cite{Espaillat_2019}.  This might be related to 
another key finding, namely that accretion rates derived from X-rays fall short
of those derived from UV/optical tracers by an order of magnitude
\citep{Curran_2011}. Conceptually, the FUV and the soft X-rays should be the
best tracers for the accretion rate, because both the FUV and the X-ray emission is
produced very close to the accretion shock where the energy budget is dominated
by the kinetic energy of the infalling mass. The X-ray emission is formed right
after the material passes through the shock front. Models successfully
reproduce line ratios in the spectrum \cite{Lamzin_2004,Guenther_2007}, although
they do not describe the full temperature range of the shock accurately
\cite{Brickhouse_2010}. The FUV emission is formed at the bottom of the
accretion column and in the gas around it
\cite{Calvet_1998,Robinson_2019} and yet, X-ray and FUV variability
seem uncorrelated or even anti-correlated \cite{Schneider_2018}. This finding 
clearly points to a fundamental problem in our
current understanding of the accretion shock.

What is the impact of this variability on the FUV emission from the disk?
Section~\ref{sect:bump} discusses the 1600\,\AA{} bump, which is most
likely due to emission from excited H$_2$, i.e.\ it originates in the disk. In
a sample of five CTTSs observed five times each with HST/STIS, the flux of this
feature correlates with the accretion luminosity determined from the UV
continuum \cite{Espaillat_2019}, but not with the X-ray flux. This is
consistent with the idea that the H$_2$ is excited by the stellar Ly$\alpha$ 
emission.

Another disk related feature that has a great impact on the FUV appearance 
of CTTSs are long-lasting dimming events such as the one observed 
for AA Tau \citep{Bouvier_2013}. During such dimming events, the 
optical brightness of the system decreases by a few magnitudes for months 
to years, i.e., they are significantly dimmer for a longer time-period than 
typically observed. While the precise cause for the dimming is not clear, it is likely
caused by changes in the inner disk structure, precisely the disk region traced
by UV-H$_2$ emission (see sect.~\ref{sect:H2disk}). Therefore, comparison of 
the UV-H$_2$ emission before and during a dimming event provides key information
on the disk regions that take part in the dimming event. \citet{Schneider_2015}
find that the observed UV-H$_2$ line profiles lack emission from regions
within 2\,au from the star while emission originating at larger radii is unaffected 
\citep[also][]{Hoadley_2015}. Furthermore, \citet{Schneider_2015} find that the 
flux ratio between the dim and bright state cannot be explained by simply adding
extra absorption for both, the continuum and the red line wings of hot ion 
lines (see Fig.~\ref{fig:AA_Tau}). These authors suggest that scattering is 
the most plausible explanation, because other scenarios require  
uncomfortable fine-tuning, 
e.g., anomalous dust requires not only a very peculiar dust size distribution, 
but also an exchange of a precise fraction of dust obscuring the system during 
the bright state with a special dust column. Lastly, the blue line wings of 
hot ion lines are only weakly affected by the dimming suggesting that they have 
strong outflow/jet contributions already in the bright state. Such dimming 
events therefore provide unique information as the central star is hidden behind 
a natural coronograph. Unfortunately, these dimming events are rare and 
unpredictable so that following each new discovery would be highly valuable.

Young stars also exhibit much longer lasting and
more consequential outbursts, where the disk structure changes
dramatically and the accretion rate increases by several orders of
magnitude. The longest and brightest of these outbursts are called ``FU Ori
outbursts'', and it can take decades or centuries until the
accretion rate settles back to the previous, stable state (see review in
\cite{Reipurth_2010}). In FU~Ori outbursts,
the system is totally dominated by the accretion luminosity. The UV spectrum
looks very different from a CTTS: It appears to be a superposition of a hot
(9000~K), 
geometrically thick inner disk and a cooler outer component (5000~K), dominated
by a dense forest of absorption lines from disk outflows
\cite{Kenyon_1989,Kravtsova_2007}.

In outflows and jets, we see temporal variability in the mass launching.
In the jets, distinct emission regions, called ``knots'', can be seen,
which are either blobs of mass launched from the star over periods of
years or shock fronts traveling along the slower moving material of
the jet. One of the best examples of this in the FUV is the jet of
HD~163296, a Herbig Ae star. This star shares many characteristics with
CTTSs, but because it is more massive, and thus brighter, the accretion and 
jet features can be observed with comparably higher signal-to-noise ratios.   
Repeated observations of Ly$\alpha$ emission in the jets shows
that knots observed at one period faint away and new knots emerge from the
source of the jet \cite{Devine_2000,Guenther_2013}. Similar observations have
been made for CTTSs in other bands (e.g.\ in DG~Tau four knots have been
tracked for two decades in the optical, X-ray and radio \cite{Rodriguez_2012}),
so it is likely that a time-series of Ly$\alpha$ images would show the flow of 
the knots along the jets in CTTSs, too, if they were observed with a 
sufficiently sensitive instrument.

\section{Conclusions and Outlook} 
Young stars are strong UV emitters compared to their older siblings.
This emission is mostly due to processes that are intrinsic to star 
formation, namely, protostellar mass accretion, enhanced stellar activity, and emission from surrounding disks and outflows. In addition, the UV radiation field controls the 
conditions in the planet forming region of protoplanetary disks 
and is thought to drive or contribute to the eventual dispersal 
of the disks. Therefore, UV observations are critical to 
painting the complete picture of star- and planet-formation.   

In this review, we have discussed, first, that the 
chromospheric and transition region emission is particularly
strong in CTTSs due to enhanced magnetic activity as young stars 
are generally fast rotators. Second, we showed that the accretion 
processes releases most 
of the energy in the UV range as continuum and Ly$\alpha$
emission. Accretion also causes strong emission lines from hot 
ions (e.g. C~{\sc iv}), which provide detailed accretion kinematics 
not available in other wavelength ranges. 
Their study suggests that accretion funnels are not homogeneous structures
but rather stratified in density so that optical depth effects are 
important and probably only a fraction of the accretion funnel is seen 
depending on inclination angle. Third, UV observations provide important,
complementary diagnostics for the structure and evolution of 
protoplanetary disks and, fourth, jets and outflows are also seen in 
the UV regime where new observations revealed jet components not seen
in traditional jet tracers, namely a hot, stationary jet
component close to the central star. Fifth, young stars are highly
variable and monitoring campaigns combining X-ray, UV, and optical data
are beginning to disentangle how the individual features are causally
related. Despite all this success, there is no theory that can 
self-consistently explain all features. The coming years will 
see increasingly sophisticated models, incorporating,
e.g., a more detailed treatment of the radiative 
effects of the accretion funnel including the reprocessing of the 
primary X-ray photons.

While most of the contributions to the system's FUV emission have now 
been identified, there are still potentially large uncertainties that 
need to be addressed by future observations. First, scattering may be 
important in more obscured sources or sources seen under 
high inclination angles \cite{Schneider_2015}; it is currently 
unknown how much scattering contributes to the observed UV spectrum of a CTTS. This is important, because
the de-reddening corrections that apply to the central star may not 
apply to other FUV emission components. 
High sensitivity two-dimensional imaging spectroscopy (e.g., an IFU 
or MSA-enabled instrument)  would allow us to simultaneously measure the inner 
disk region as a function of wavelength, spatial location, and velocity; 
allowing us to quantify the relative importance of scattering and the
temporal variability of different disk  components individually.  
Second, jets can contribute to
the FUV emission and, because their emission is located close to the 
star, their emission can be mistaken as stellar; note that there are 
not necessarily kinematic signatures like highly blue-shifted emission
as the blue- and red-shifted jet lobe may both emit resulting in a 
net velocity shift close to zero. Future high resolution imaging spectroscopy 
observations are needed to 
constrain the importance of scattering and jet emission in ``typical''
samples of CTTSs. 

The next major step for understanding the UV emission of CTTSs will be the 
Hubble Space Telescope's ULLYSES program. The ULLYSES program will dedicate
$\sim$~500 orbits to the study of low-mass star formation in the UV.
ULLYSES will include a spectral atlas of $\sim$~60 new sources and 
a dedicated, multi-wavelength monitoring program for high-priority targets. 
On intermediate time scales, however, 
neither ESA nor NASA are committed to build a space mission with
dedicated UV instrumentation within the next decade. The Russian Spektr-UV mission
may be the next dedicated UV mission and is currently scheduled to launch in the mid-2020s. With a 1.7\,m mirror, it will be somewhat comparable to HST; also in it's spectroscopic and imaging capabilities (low- and high-resolution UV spectroscopy).  The next opportunity to 
launch a large space mission with sensitive UV spectroscopic capabilities may be the 
LUVOIR mission concept \citep{2017_LUVOIR,LUVOIR_2018}, to be launched in 
the late 2030s. In fact, LUVOIR's science goals
specifically include the questions of low-mass star formation that we 
discussed in the review.

\vspace{6pt} 



\authorcontributions{writing--review and editing, P.C.S., H.M.G, K.F.; visualization, P.C.S.}

\funding{This research was funded by DLR grant number 50 OR 1907. K.F. acknowledges support for this work from HST GO programs 14604, 14703, and 15070. The authors also express their gratitude to N. Arulanantham for providing information on the H$_2$ emission properties of CTTSs.}


\conflictsofinterest{The authors declare no conflict of interest.} 

%

\appendixtitles{no} 
%

\reftitle{References}


\externalbibliography{yes}
\bibliography{uv}





\end{document}